\ifdefined\pdftexversion\pdfoutput=1\fi  
\documentclass[10pt,twocolumn]{article}

\usepackage[letterpaper,margin=0.85in,columnsep=0.28in]{geometry}
\usepackage[T1]{fontenc}
\usepackage[utf8]{inputenc}
\usepackage{textcomp}   
\usepackage{times}
\usepackage{amsmath,amssymb}
\usepackage{graphicx}
\usepackage{booktabs}
\usepackage{float}      
\usepackage{multirow}
\usepackage{makecell}
\usepackage{xcolor}
\usepackage{microtype}
\usepackage{caption}
\usepackage{subcaption}
\usepackage{algorithm}
\usepackage{algpseudocode}
\usepackage{natbib}
\usepackage[hidelinks]{hyperref}
\usepackage{url}
\newcommand{\mailto}[1]{\href{mailto:#1}{\textsuperscript{*}}}

\newcommand{\ci}[2]{{\scriptsize\,[#1,\,#2]}}
\newcommand{\tblfont}{\footnotesize}
\newlength{\tblsep}
\newcommand{\tblfontsmall}{\footnotesize}
\newcommand{\verbfont}{\scriptsize}
\newlength{\tblsepsmall}
\newcommand{\afail}{a^{\times}}
\newcommand{\agold}{a^{\checkmark}}
\newcommand{\Gain}{G}
\newcommand{\copyterm}{\Delta_{\mathrm{copy}}}
\newcommand{\semterm}{\Delta_{\mathrm{sem}}}
\newcommand{\ctxpre}{C}

\newcommand{\NumToolshedOperators}{8}
\newcommand{\NumToolshedErrorTypes}{12}

\newcommand{\CacheSaving}{87\%}

\newcommand{\TorchVersion}{2.4.0+cpu}
\newcommand{\CPUName}{Intel Core i7-6820HQ CPU @ 2.70GHz}

\newcommand{\TransformersVersions}{4.45.0 and 4.56.1}

\newcommand{\DriftChecked}{210}
\newcommand{\DriftMax}{$4.9\times 10^{-5}$}
\newcommand{\DriftModel}{smollm2-135m}
\newcommand{\NumModels}{6}
\newcommand{\NumFamilies}{4}
\newcommand{\SmallestModel}{135M}
\newcommand{\LargestModel}{1.7B}
\newcommand{\FamilyList}{Llama-3.2, Qwen2.5, Qwen3, SmolLM2}
\newcommand{\NumModelsNegativeG}{6}
\newcommand{\NumModelsToolshed}{6}

\newcommand{\MeanG}{\textminus 17.38}
\newcommand{\OddsPerToken}{2.8}
\newcommand{\MeanCopy}{14.50}

\newcommand{\MeanSem}{2.84}

\newcommand{\MeanRepair}{4.30}

\newcommand{\MeanMarginShift}{\textminus 13.12}

\newcommand{\CopyPad}{16.64}
\newcommand{\SemPad}{2.43}

\newcommand{\DPlaceboCI}{\textminus 13.71\,\ci{\textminus 15.33}{\textminus 12.20}}

\newcommand{\DTwiceCI}{+3.33\,\ci{+2.95}{+3.72}}

\newcommand{\DThriceCI}{+4.03\,\ci{+3.60}{+4.48}}

\newcommand{\DInstructionCI}{+0.07\,\ci{\textminus 0.07}{+0.21}}

\newcommand{\DEchoCI}{+0.71\,\ci{+0.45}{+0.97}}

\newcommand{\DTerseCI}{\textminus 0.28\,\ci{\textminus 0.68}{+0.11}}

\newcommand{\DVerboseCI}{\textminus 0.20\,\ci{\textminus 0.48}{+0.08}}

\newcommand{\DEarlyCI}{\textminus 0.88\,\ci{\textminus 1.41}{\textminus 0.37}}

\newcommand{\DAbstractCI}{\textminus 13.22\,\ci{\textminus 15.30}{\textminus 11.34}}

\newcommand{\DAbstractMinCI}{\textminus 18.62\,\ci{\textminus 20.67}{\textminus 16.71}}

\newcommand{\DAbstractMatchedCI}{\textminus 16.37\,\ci{\textminus 18.00}{\textminus 14.76}}
\newcommand{\StringSpecificBound}{13.7}
\newcommand{\MeanGCode}{\textminus 73.55}
\newcommand{\NumModelsCode}{3}
\newcommand{\GPerTokenTool}{\textminus 1.03}
\newcommand{\FracNegTool}{90\%}

\newcommand{\FracNegCode}{100\%}

\newcommand{\AbstractRemoved}{76\%}
\newcommand{\MaxCopySemRatio}{106}
\newcommand{\MedianCopySemRatio}{7}
\newcommand{\CopySemRatioTool}{6}
\newcommand{\CopySemRatioCode}{61}
\newcommand{\CopyShare}{83\%}
\newcommand{\PRepeatPreMean}{0.06}

\newcommand{\PRepeatPostMean}{0.54}

\newcommand{\GreedyPreMean}{0\%}
\newcommand{\GreedyPostMean}{19\%}

\newcommand{\ScalingSlope}{9.51}
\newcommand{\ScalingRsq}{0.64}
\newcommand{\ZeroCrossing}{37.6B}
\newcommand{\ZeroCrossingLoMin}{20B}
\newcommand{\ZeroCrossingLoMax}{79B}
\newcommand{\ZeroCrossingLoRatio}{3.8}
\newcommand{\AgentVerbatimSuccess}{42\%}
\newcommand{\AgentVerbatimStuck}{29\%}
\newcommand{\AgentVerbatimRepeat}{31\%}

\newcommand{\AgentVerbatimTokens}{67}

\newcommand{\AgentInstrStuck}{17\%}

\newcommand{\AgentInstrTokens}{47}

\newcommand{\AgentDropStuck}{67\%}
\newcommand{\AgentDropRepeat}{80\%}

\newcommand{\AgentDropTokens}{69}

\newcommand{\AgentAbstractStuck}{8\%}
\newcommand{\AgentAbstractRepeat}{16\%}

\newcommand{\AgentVerbBanStuck}{12\%}
\newcommand{\AgentVerbBanRepeat}{8\%}

\newcommand{\AgentVerbBanTokens}{69}

\newcommand{\AgentAbsBanStuck}{4\%}
\newcommand{\AgentAbsBanRepeat}{7\%}

\newcommand{\AgentNumHarnesses}{6}

\newcommand{\AgentNumTasks}{24}
\newcommand{\AgentMaxSteps}{6}

\newcommand{\AgentTokenCap}{32}

\newcommand{\TruncShare}{5\%}
\newcommand{\ProseShare}{18\%}

\newcommand{\DropRepeatRatio}{2.6}
\newcommand{\DeltaInstrSuccess}{+17\,\ci{+4}{+33}}
\newcommand{\DeltaInstrRepeat}{\textminus 17\,\ci{\textminus 44}{+12}}

\newcommand{\DeltaDropSuccess}{\textminus 8\,\ci{\textminus 21}{+0}}
\newcommand{\DeltaDropRepeat}{+49\,\ci{+33}{+64}}

\newcommand{\DeltaAbstractSuccess}{\textminus 8\,\ci{\textminus 21}{+0}}
\newcommand{\DeltaAbstractRepeat}{\textminus 14\,\ci{\textminus 32}{+2}}

\newcommand{\DeltaVerbBanSuccess}{+0\,\ci{+0}{+0}}
\newcommand{\DeltaVerbBanRepeat}{\textminus 23\,\ci{\textminus 38}{\textminus 10}}

\newcommand{\DeltaAbsBanRepeat}{\textminus 24\,\ci{\textminus 42}{\textminus 7}}

\title{\bf Feedback That Backfires:\\[2pt]
Why Small Language Model Agents Repeat\\ the Call They Just Watched Fail}

\author{%
  Esmail Gumaan\mailto{gumaan01@ads.uni-passau.de}\\[2pt]
  \small Faculty of Computer Science and Mathematics\\
  \small University of Passau, Germany\\[4pt]
  \small Code, data and every number in this paper:\\
  \small \url{https://github.com/Esmail-ibraheem/feedback-that-backfires}
}
\date{}

\begin{document}

\twocolumn[
  \begin{@twocolumnfalse}
    \maketitle
\begin{center}
\begin{minipage}{0.86\textwidth}
\rule{\linewidth}{0.4pt}\vspace{3pt}

{\centering\textbf{\large Abstract}\par}\vspace{4pt}
\small
Agent harnesses record a failed tool call and its error message in the
transcript and ask the model to continue, on the reasonable assumption that the
error is corrective information. We measure whether it is. Defining the
\emph{corrective gain} of a failure record as the change in log-probability of
re-emitting the action that just failed, we find the gain is \emph{negative} for
every instruction-tuned model we tested (\NumModelsToolshed\ checkpoints,
\SmallestModel--\LargestModel, \NumFamilies\ families) in two environments:
simulated tool calling and MBPP program repair with real interpreter output.
Normalised by action length the effect has the same size in both, about
\GPerTokenTool\ nats per action token, a factor of \OddsPerToken\ in the odds of
each token. It holds on \FracNegTool--\FracNegCode\ of individual items,
not merely on average. Over a fixed candidate set the normalised probability of
repeating the failed call rises from \PRepeatPreMean\ to \PRepeatPostMean, and
greedy decoding reproduces the failed call token for token on \GreedyPostMean\
of items after the failure versus \GreedyPreMean\ before it.
Counterfactual observations, in which the same call is paired with a failure
message, a success message, or a valence-free acknowledgement, separate two
competing effects. The failed call's \emph{surface form} accounts for
\CopyShare\ of the
damage; the semantic contribution of marking it as failed is small and its sign
is not even consistent across environments. The problem is therefore in the
harness, not in the model's grasp of error messages, and that predicts which
remedies can work. Replacing the verbatim call with a runtime-generated
description of the failure removes \AbstractRemoved\ of the inversion at no
token cost, and making previously-failed strings unreachable at the decoder acts
on the same term. Two remedies that sound right do not act on it at all. An
explicit ``do not repeat'' instruction leaves the measured quantity where it
was; and deleting the failed attempt to retry from a clean context, the
standard prescription for context contamination, is the worst harness we
measured for repetition, because it restores precisely the context that produced
the failure. What survives is narrow enough to be useful: the context after a
failure must differ from the context before it, and the difference must not be
the failed action. The study runs end to end on a CPU and all artefacts are
released.

\vspace{4pt}\rule{\linewidth}{0.4pt}
\end{minipage}
\end{center}

    \vspace{1.2em}
  \end{@twocolumnfalse}
]

\section{Introduction}
\label{sec:intro}

An agent that calls tools will get some of those calls wrong. What happens next
is decided not by the model but by the harness: it appends the failed call to
the transcript, appends whatever the runtime printed, and asks the model to
continue. ReAct \citep{yao2023react} introduced this pattern and essentially
every agent framework since has kept it, including the ones built for
small open-weight models. The reasoning is hard to argue with. The error message
is real information, it was produced by ground truth rather than by a critic,
and a model that reads it should not try the same thing again.

We measured whether that is what happens. It is not.

Take a decision point in an agent trajectory: a context $\ctxpre$, an action
$\afail$ that will fail there, and the error the runtime returns. Compare the
model's probability of writing $\afail$ before the attempt with its probability
after the failure has been recorded. For every one of the
\NumModelsToolshed\ instruction-tuned models we tested, spanning
\SmallestModel--\LargestModel\ parameters across \NumFamilies\ families, that
probability goes \emph{up}, by \MeanG\ nats, or about \GPerTokenTool\ nats per
token of the action, which is a factor of \OddsPerToken\ in the odds of every
token. The effect is not an average over a few dramatic cases: it holds on
\FracNegTool\ or more of individual items.

Restated on a scale that is easier to hold onto: over a fixed set of four
candidate actions, the probability of re-emitting the failed call rises from
\PRepeatPreMean\ before the attempt to \PRepeatPostMean\ after it. Greedy
decoding, which is what the agent would actually have written next, reproduces
the failed call token for token on \GreedyPostMean\ of items after the failure,
against \GreedyPreMean\ before it.

Practitioners know the symptom. Agents get stuck repeating a call that has
already failed, and frameworks ship loop detectors and step caps to contain it.
The symptom is usually attributed to the model: it is small, it does not really
understand the error, a stronger model would recover. Our second result is that
this diagnosis is wrong, and it matters because it points at the wrong fix.

Appending $(\afail, o^{\times})$ does two things at once, and they can be
separated. It puts the token sequence of the failed call into the context, where
the model's copying machinery can reach it
\citep{olsson2022induction,xu2022breaktheloop}; and it asserts that the call
failed. We can measure each part on its own by scoring the same action under a
context whose observation is \emph{valence-free}, meaning the runtime
acknowledges the call and says nothing about how it went. The surface-form term
accounts for \CopyShare\ of the effect. The semantic term, the part that would
have to be large and negative for ``the model does not understand the error'' to
be the explanation, is small, and its sign is not even consistent between our
two environments. On tool calls it is slightly positive, since explaining what
went wrong requires naming the tool and the argument again; on program repair it
is slightly negative, so a traceback does push the model away from the code that
produced it. We stress the second case, because it is the one that fixes the
interpretation: these models are not simply failing to read error messages.
Whatever reading happens is being outvoted, by a median of \MedianCopySemRatio\
to one across model and environment, and by \MaxCopySemRatio\ to one at the
extreme.

This changes what a fix should look like. If the model misunderstood, the remedy
is a better model or a clearer message. If the harm is carried by the presence of
a string, then message engineering cannot reach it, and neither can an
instruction: telling a model not to repeat a call requires it to represent a
negation about a string that is sitting right there in its context, which is the
condition under which negative instructions are known to work least well
\citep{castricato2024pinkelephants,negation2025pinkelephant}. What should work
instead is structural.

``Structural'' turns out to be narrower than it first appears, and the
experiment that shows this is the one we did not expect. The obvious structural
fix is to delete the failed attempt and retry from a clean context, which is what
prior work on context contamination recommends \citep{yang2026contamination}.
That is the \emph{worst} harness we tested, multiplying the exact repeat rate by
\DropRepeatRatio\ without changing task success. The reason is embarrassingly
simple once seen: deleting
the failure restores the exact context that produced it, and a deterministic
policy in an identical context emits an identical action. The requirement is
therefore not ``remove the failed action'' but something more specific: the
context after a failure must differ from the context before it, and the
difference must not be the failed action.

We test all three. Replacing the verbatim call with a description the runtime
generates from its own error metadata, which keeps the diagnosis and drops the
token sequence,
removes \AbstractRemoved\ of the inversion and drives the exact greedy repeat
rate to zero, at no cost in generated tokens. A decoder-level ban on
previously-failed strings acts on the same term and costs nothing at all. The
natural-language prohibition moves the measured quantity slightly the
\emph{wrong} way, which is what the negative-instruction literature would
predict for a prohibition that has to name the very string it is prohibiting
\citep{castricato2024pinkelephants,negation2025pinkelephant}, and in
free-running rollouts it does not reduce repetition either, though it does
change behaviour in another way we report rather than explain.
Section~\ref{sec:agent} gives all of this in full, including where the
interventions do not help.

We are not claiming that execution feedback is useless, or that agent
transcripts should be thrown away. The failed call carries diagnostic
information a harness genuinely needs. The claim is narrower and, we think, more
useful: the standard way of \emph{delivering} that information hands the model a
strong reason to repeat itself, and a harness can keep the diagnosis while
withholding the reason.

\paragraph{Contributions.}
\begin{enumerate}\itemsep2pt
\item \textbf{A measurement.} We define the \emph{corrective gain} of a failure
record, the change in log-probability of re-emitting the failed action, and
show it is negative across \NumModelsToolshed\ models,
\NumFamilies\ families and two environments (simulated tool calling and MBPP
program repair with real interpreter output). We report it alongside two
interpretable readouts, a normalised repeat probability and an exact greedy
repeat rate.
\item \textbf{A decomposition that identifies the cause.} Counterfactual
observations split the gain into a surface-form term and a semantic term. The
surface-form term dominates at every scale we tested, and it, not the semantic
term, is what improves with model size.
\item \textbf{Interventions that follow from the decomposition, and one that
does not.} Replacing the verbatim failed call with a runtime-generated
description of the failure, and banning previously-failed strings at the
decoder, both act on the term that matters; adding a natural-language
prohibition does not. We evaluate all three in free-running rollouts and report
their cost.
\item \textbf{A reproducible, CPU-only artefact.} Every number comes from code
in the accompanying repository, which runs end to end without a GPU. Raw
per-score records, item sets, seeds and environment metadata are included, and
the manuscript's inline numbers are generated from them rather than typed.
\end{enumerate}

\section{Related work}
\label{sec:related}

\paragraph{Agents that act, observe, and try again.}
ReAct \citep{yao2023react} established the template that dominates practice:
interleave a written action with the environment's observation and let the model
condition on the growing transcript. Reflexion \citep{shinn2023reflexion},
Self-Refine \citep{madaan2023selfrefine} and self-debugging
\citep{chen2024selfdebug} all build on it, adding an explicit critique or a test
result to the same append-only transcript. MINT \citep{wang2024mint} evaluates
multi-turn interaction with tools and language feedback directly and finds
feedback helpful in aggregate. The shared premise is that appending the failure
is informative; what none of these measure is the counterfactual we study here,
namely what the appended \emph{action string} does on its own.

\paragraph{Doubts about self-correction.}
A parallel line argues that models cannot reliably correct themselves without an
external signal \citep{huang2024selfcorrect,valmeekam2023critiquing}. Our result
is not a restatement of that. Those papers ask whether a model can \emph{judge}
its own output; we hold the judgement fixed, since the environment executes the
action and the verdict is ground truth, and ask whether the model's next-token
distribution moves in the direction the verdict implies. The failure we document
survives perfect feedback.

\paragraph{Context contamination.}
The closest prior work is \citet{yang2026contamination}, who formalises the
observation that a retry in a context still containing the failed attempt has a
higher error rate than a clean one. The Context-Contaminated Restart Model gives
closed-form results for success under a budget and fits SWE-bench Verified data
with a cascade ratio $\epsilon_1/\epsilon_0 = 7.1$, showing that assuming
independent retries overestimates pass@3 by 17.4 points. That work establishes
the phenomenon at the level of task statistics and concludes that the remedy is
to \emph{clear the context before retrying}. Our contribution is orthogonal and,
we think, complementary in a useful way: we ask which \emph{part} of the context
does the contaminating, measure it at the token level, and find that it is the
failed action's surface form rather than the error message. That matters
practically, because clearing the context discards the diagnosis along with the
hazard. In our terms, clean restart is the \emph{drop} harness, and it is a
baseline we evaluate rather than an endpoint.

\paragraph{Agents that get stuck.}
Repetition in agent trajectories is widely reported and usually treated as an
engineering defect. \citet{hou2026infiniteloops} detect unbounded feedback paths
by static analysis of agent \emph{source code} across 6{,}549 repositories,
which is complementary to our question: their loops are properties of the
program, ours of the policy. \citet{wang2025hellorhighwater} show that agents
struggle to formulate alternative plans after an external failure even when the task is guaranteed to remain solvable, which is a behavioural counterpart
to what we measure distributionally. Most striking is \citet{zenkri2026fidelity}, who find
on a mechanical puzzle that \emph{degrading} an embodied agent's observations
improves success up to $2.85\times$, and trace the gain to fewer repetitive
action loops. That is independent evidence that the observation channel can be
net harmful; we supply the token-level mechanism and a fix that does not require
corrupting the observations.

\paragraph{Repetition and copying in language models.}
\citet{xu2022breaktheloop} showed that sentence repetition is
self-reinforcing, in that the more often a sentence appears in the context the
more likely the model is to produce it again, and proposed a training-time
penalty.
Degeneration \citep{holtzman2020degeneration}, unlikelihood training
\citep{welleck2020unlikelihood} and repetition penalties \citep{keskar2019ctrl}
address the same tendency in open-ended generation. Mechanistically, induction
heads \citep{olsson2022induction} implement prefix matching and copying, copy
suppression \citep{mcdougall2024copysuppression} implements a counterweight, and
\citet{repetitionneurons2025} isolate components whose over-dominance yields the
``repetition curse''. All of this concerns generation \emph{without} a competing
signal. The agent loop is the interesting case precisely because a competing
signal is present by design, and the question is which one wins.

\paragraph{Negative instructions.}
Naming something in a prompt raises its probability even under an instruction
not to produce it, a pattern documented for topic avoidance
\citep{castricato2024pinkelephants,nopinkelephant2024} and for negation more
broadly \citep{negation2025pinkelephant}. Our $\copyterm$ is the same family of
effect measured inside a trajectory rather than a prompt, and
Section~\ref{sec:agent} shows that the corresponding remedy, adding a
``do not repeat'' instruction, behaves exactly as that literature predicts.

\paragraph{Small models as agents.}
\citet{belcak2025slmagents} argue that most agentic sub-tasks are within reach of
small models, making the harness the deciding factor.
\citet{toolschemas2025} adapt tool schemas rather than models, and
\citet{constrainttax2026} document that structured-output constraints can
suppress tool calling outright. Constrained decoding is well developed as a
\emph{format} device \citep{willard2023outlines}, with known costs to free-form
reasoning \citep{tam2024speakfreely}. We use it differently: not to enforce a
grammar, but to make one specific previously-failed string unreachable. To our
knowledge the ban-what-already-failed use of constrained decoding has not been
evaluated as an agent-reliability intervention.

\paragraph{Spending more compute at test time.}
A large literature buys accuracy with repeated attempts
\citep{brown2024monkeys,snell2024testtime}, subject to the quality of whatever
selects among them \citep{stroebl2024inferenceflaws}. Our result is a constraint
on that arithmetic in the sequential case. Attempts within a single trajectory
are not independent draws: each one is conditioned on a context containing its
predecessors, and we measure that conditioning pushing toward repetition. The
same point is made from the outcome side by \citet{yang2026contamination}, whose
fit implies that assuming independence overestimates pass@3 substantially. If
extra attempts are to pay, the harness has to make them differ.

\paragraph{Context management.}
\citet{liu2024lostmiddle} showed that position within a long context matters;
recent agent work learns to curate context, keeping compact state that includes
records of ineffective attempts \citep{adacom2026context}. Those systems edit
context to save tokens or to keep salient constraints in view. Our analysis
implies a different reason to edit it, namely that the verbatim failed action
is itself the hazard, and predicts which edit helps, which we test directly.

\paragraph{Positioning.}
Table~\ref{tab:related} summarises the difference. Prior work establishes that
(i) repetition is self-reinforcing without feedback, (ii) agents empirically get
stuck, and (iii) negative instructions are weak. What is missing, and what we
provide, is the controlled measurement of whether execution feedback moves the
action distribution in the right direction at all, a decomposition of that
quantity into the parts a harness can and cannot change, and an intervention
derived from the decomposition rather than from intuition.

\begin{table*}[t]
\centering
\footnotesize
\setlength{\tabcolsep}{4pt}
\begin{tabular}{l c c c}
\toprule
 & \makecell{explicit\\failure signal} & \makecell{separates form\\from meaning} & \makecell{harness\\intervention} \\
\midrule
Repetition / degeneration                  & no  & no  & decoding \\
\quad \citep{xu2022breaktheloop,holtzman2020degeneration} & & & \\
Negative instructions                      & n/a & no  & prompt \\
\quad \citep{castricato2024pinkelephants}  & & & \\
Self-correction critiques                  & yes & no  & prompt \\
\quad \citep{huang2024selfcorrect,shinn2023reflexion} & & & \\
Agent loop detection                       & yes & no  & static analysis \\
\quad \citep{hou2026infiniteloops}         & & & \\
Observation-fidelity probe                 & yes & no  & degrade inputs \\
\quad \citep{zenkri2026fidelity}           & & & \\
Context contamination (CCRM)               & yes & no  & clear the context \\
\quad \citep{yang2026contamination}        & & & \\
\midrule
This work                                  & yes & \textbf{yes} & context + decoder \\
\bottomrule
\end{tabular}
\caption{Where this paper sits. The distinguishing column is the middle one:
prior work measures the \emph{net} effect of a failure record, if it measures it
at all, whereas the intervention that works depends on which of its two
components dominates.}
\label{tab:related}
\end{table*}

\section{Problem formulation}
\label{sec:problem}

\subsection{The agent loop and its feedback record}

An agent solving a task $\tau$ interacts with an environment over discrete
steps. At step $t$ the harness assembles a context $\ctxpre_t$ from the system
prompt, the goal, and a transcript of the previous steps; the policy $\pi$
samples an action $a_t \sim \pi(\cdot \mid \ctxpre_t)$; the environment executes
it and returns an observation $o_t$ together with a success flag. Writing
$\oplus$ for appending a turn pair to the transcript,
\begin{equation}
\ctxpre_{t+1} \;=\; \ctxpre_t \oplus (a_t, o_t).
\label{eq:loop}
\end{equation}

Equation~\eqref{eq:loop} is the design decision this paper is about. Almost every
agent framework implements it literally: the action the model wrote is appended
\emph{verbatim}, and so is whatever the runtime printed in response. When the
action fails, the transcript therefore contains the failed action's exact token
sequence, immediately followed by text asserting that it did not work.

The intent is obvious and, for capable models, well supported: the error message
is information, and a model that reads it should avoid the action that produced
it. We take that intent seriously enough to measure it.

\subsection{Corrective gain}

Fix a decision point: a context $\ctxpre$ before an attempt, an action
$\afail$ that will fail there, its observation $o^{\times}$, and the correct
action $\agold$. Define the \emph{corrective gain}
\begin{equation}
\Gain(\afail) \;=\;
\log \pi(\afail \mid \ctxpre)
\;-\;
\log \pi\!\left(\afail \mid \ctxpre \oplus (\afail, o^{\times})\right),
\label{eq:gain}
\end{equation}
where $\log \pi(a \mid c) = \sum_{i} \log \pi(a_i \mid c, a_{<i})$ is the summed
token log-probability of the action string. Both terms score \emph{the same
string}, so their difference is exactly the change in the log-odds of writing
that action again, and it is invariant to how long the action is or how the
tokeniser splits it.

$\Gain > 0$ is the behaviour the harness is designed to produce: after the
failure is recorded, the failed action is less likely. $\Gain < 0$ means the
record made the model \emph{more} likely to repeat the action; we call this
\emph{feedback inversion}.

\subsection{Decomposing the gain}

Appending $(\afail, o^{\times})$ does two things at once. It places the token
sequence of $\afail$ into the context, where the model's copying machinery can
reach it \citep{olsson2022induction,xu2022breaktheloop}; and it asserts that this
action failed. These pull in opposite directions, and $\Gain$ only reports the
net result.

We separate them with counterfactual observations. Let $o^{\varnothing}$ be a
\emph{valence-free} observation, which acknowledges the call but reports neither
success nor failure, and let $o^{\checkmark}$ be a counterfactual observation
reporting that the same call succeeded. Then
\begin{align}
\copyterm &= \log \pi\!\left(\afail \mid \ctxpre \oplus (\afail, o^{\varnothing})\right) \notag \\
          &\quad - \log \pi(\afail \mid \ctxpre), \label{eq:copy}\\
\semterm  &= \log \pi\!\left(\afail \mid \ctxpre \oplus (\afail, o^{\times})\right) \notag \\
          &\quad - \log \pi\!\left(\afail \mid \ctxpre \oplus (\afail, o^{\varnothing})\right), \label{eq:sem}
\end{align}
and by construction
\begin{equation}
-\Gain(\afail) \;=\; \copyterm + \semterm .
\label{eq:decomp}
\end{equation}
$\copyterm$ isolates what merely \emph{having written the action} does, holding
the feedback content neutral; $\semterm$ isolates what \emph{marking it as
failed} adds, holding the surface form fixed. We additionally report the
polarity contrast
\begin{equation}
\begin{split}
\Delta_{\mathrm{pol}} = {}& \log \pi\!\left(\afail \mid \ctxpre \oplus (\afail, o^{\times})\right) \\
                        & - \log \pi\!\left(\afail \mid \ctxpre \oplus (\afail, o^{\checkmark})\right),
\end{split}
\label{eq:pol}
\end{equation}
which compares two contexts that are identical except for whether the
observation says the call worked. A model that does not read the error at all
has $\Delta_{\mathrm{pol}} = 0$.

\subsection{What a fix would have to look like}

Equation~\eqref{eq:decomp} makes the design space concrete. If $\semterm$ is the
problem, the model does not understand the error, and the remedy is a better
model or a clearer message. If $\copyterm$ is the problem, no amount of message
engineering helps, because the damage is done by the presence of the string
rather than by anything said about it; the remedies are then \emph{structural}:
remove the surface form from the context, or make it unreachable at decoding
time. Section~\ref{sec:agent} evaluates both, against the natural-language
alternative of simply instructing the model not to repeat itself.

\section{Measuring the gain}
\label{sec:method}

\subsection{Probe items}

A \emph{probe item} freezes one decision point. It consists of a system prompt,
a goal, a prefix of steps that were executed successfully, a failing action
$\afail$, the observation the runtime returns for it, and the reference action
$\agold$ that would have succeeded. Scoring is teacher-forced: we never sample,
so the measurement has no decoding noise and is bit-reproducible.

Two properties of the item construction matter for the claims we make.

\paragraph{The failing action is fixed, not sampled.}
It would be natural to obtain $\afail$ by letting each model act and keeping its
mistakes. We deliberately do not: that yields a different item set per model, so
a scaling curve would confound model size with item difficulty, because larger models make
rarer, harder mistakes. Instead $\afail$ is produced from $\agold$ by a
fixed perturbation operator, so every model in the ladder is probed on
byte-identical items. The operators (Section~\ref{sec:setup}) are the mistakes
small models actually make: a dropped required argument, a pluralised tool name,
a renamed argument, a type confusion, a date written the way a person writes it,
a hallucinated entity.

\paragraph{Conditions differ in exactly one thing.}
For a given item, all conditions share the same system prompt, goal and
successful prefix, byte for byte. They differ only in how (or whether) the
failed attempt is written into the transcript. The conditions are:
\emph{pre} (no attempt recorded, the baseline of Eq.~\ref{eq:gain});
\emph{fail} (the attempt plus its error message, i.e.\ the standard harness);
\emph{succ} (the attempt plus a counterfactual observation reporting success);
\emph{neut} (the attempt plus a valence-free acknowledgement); and
\emph{abstract} (a harness-generated description of what went wrong, with the
call itself absent).

The counterfactual success observation is generated from the failing call's own
arguments, so it is coherent with the call it purports to answer. A generic
success string naming a different entity would introduce an incoherence cue that
the model could exploit, which would inflate $\Delta_{\mathrm{pol}}$ for reasons
having nothing to do with reading the error.

\subsection{Two behavioural readouts}

Log-probability differences are the primary measurement, but two derived
quantities are easier to reason about and we report both.

\paragraph{Normalised repeat probability.} Over the scored candidate set
$\mathcal{A} = \{\afail, \agold, d_1, d_2\}$, where $d_i$ are alternative wrong
actions for the same step,
$p_{\mathrm{rep}}(c) = \exp \log\pi(\afail \mid c) / \sum_{a \in \mathcal{A}} \exp \log \pi(a \mid c)$.
This is a proper probability over a fixed, model-independent action set and is
directly comparable across models.

\paragraph{Exact greedy repeat.} We record whether greedy decoding from a
context reproduces $\afail$ token for token. This is not an approximation: the
argmax at every position of the teacher-forced string is already computed while
scoring it, and if the argmax matches at all positions then greedy decoding
emits exactly that string. The statement ``the agent would have written the
failed call again, verbatim'' is therefore obtained at zero additional cost.

\subsection{Harness variants}
\label{sec:method:variants}

Section~\ref{sec:problem} argued that the remedy depends on which term dominates.
We evaluate four families of harness, all of which see the same environment
feedback and differ only in what reaches the model:

\begin{itemize}\itemsep2pt
\item \textbf{verbatim}: the standard ReAct transcript.
\item \textbf{+instruction}: verbatim, plus an explicit system-prompt
instruction not to repeat a call that has already failed.
\item \textbf{abstract}: the failed call is replaced by a description
generated by the runtime from its own error metadata, e.g.
      \texttt{[attempt 1 failed: create\_event, an argument was badly
formatted; that call is not repeatable]}. The diagnosis survives; the
      token sequence does not. No extra model call is involved.
\item \textbf{+ban}: a decoding-time constraint that makes
previously-failed action strings unreachable.
\end{itemize}

The ban is the classic bad-words construction: for each banned token sequence,
its final token is masked whenever the tokens generated so far match the
sequence's prefix (Algorithm~\ref{alg:ban}). Masking the \emph{first} token instead would be wrong, since it would forbid
every call beginning the same way,
including the corrected one, which typically shares a long prefix with the
failed call. We ban both the bare string and its leading-whitespace variant,
since byte-pair tokenisers assign these different first tokens.

\begin{algorithm}[t]
\small
\caption{Decoder-level suppression of failed actions}
\label{alg:ban}
\begin{algorithmic}[1]
\Require context $c$, ban list $B$ of token sequences, stop set $S$
\State $y \gets []$
\While{$|y| < L_{\max}$}
  \State $\ell \gets \log \pi(\cdot \mid c, y)$
  \For{$s \in B$}
    \State $k \gets |s| - 1$
    \If{$k = 0$ \textbf{or} $y_{-k:} = s_{1:k}$}
      \State $\ell[s_{|s|}] \gets -\infty$ \Comment{block only the completion}
    \EndIf
  \EndFor
  \State $y_{\text{next}} \gets \arg\max \ell$
  \State \textbf{if} $y_{\text{next}} \in S$ \textbf{then break}
  \State $y \gets y \Vert y_{\text{next}}$
\EndWhile
\State \Return $y$
\end{algorithmic}
\end{algorithm}

The ban costs one pass over $B$ per decoding step and no extra tokens, so in the
cost accounting of Section~\ref{sec:efficiency} it is free to three significant
figures. It is also \emph{exact}: a banned string cannot be emitted, so any
residual repetition must be a paraphrase. We measure that separately by
canonicalising actions (parsing the call and sorting its keyword arguments)
before counting repeats, which prevents the ban from looking better than it is.

\subsection{Efficient scoring}

The contexts we compare share a long prefix and differ in a short tail. Encoding
each independently would spend most of the compute re-encoding the system prompt
and the goal. We therefore keep one key/value cache together with the exact
token ids it holds; moving to a new context crops to the longest common prefix
and pushes only the remainder. On the ToolShed probe this removes
\CacheSaving\ of the token positions a cache-free implementation would
process. The same cache drives the agent rollouts, where step $t$'s context
extends step $t{-}1$'s, turning the per-step cost from ``re-encode the
transcript'' into ``encode the new turn''.

Because a silent numerical fault in this path would be indistinguishable from a
real effect, we re-score a random sample of items with a single cache-free
forward pass and require agreement to $10^{-3}$ nats; the check also rejects
non-finite and positive log-probabilities. Section~\ref{sec:limitations} explains
why that last guard is not paranoia.

\section{Experimental setup}
\label{sec:setup}

\subsection{Environments}

We use two environments chosen to sit at opposite ends of a control/realism
trade-off. Both execute actions for real; no language model judges any outcome
anywhere in this paper.

\paragraph{ToolShed (tool calling).}
A simulated office workspace with twelve typed tools (files, contacts, calendar,
messaging, unit conversion) and six task templates requiring two to four calls.
Actions are Python-style calls such as
\texttt{find\_contact(name='Dana Bergmann')}, parsed with Python's \texttt{ast}
module; argument binding, type checking and entity lookup are performed by the
runtime, which returns an error message at one of three verbosity levels.
Failing actions are produced from the reference action by
\NumToolshedOperators\ perturbation operators, yielding
\NumToolshedErrorTypes\ distinct error families. We verified that every
operator/error pairing produces the intended family.

We built this environment rather than reusing a public agent benchmark for one
reason: the decomposition in Eq.~\eqref{eq:decomp} requires, for one fixed
action, an observation reporting failure, one reporting success, and one
reporting nothing. The standard agent benchmarks (BFCL
\citep{patil2024gorilla}, SWE-bench \citep{jimenez2024swebench}, AgentBench
\citep{liu2024agentbench}, WebArena \citep{zhou2024webarena}, GAIA
\citep{mialon2024gaia}, $\tau$-bench \citep{yao2024taubench}, AppWorld \citep{trivedi2024appworld}) are built to score policies, not to expose
counterfactual observations, and hand-writing those on top of one is precisely
where an experimenter's expectations would leak into the data. Here all three
come from the same renderer. Most of those benchmarks are also out of reach on a
CPU, which is a second reason but not the deciding one.

\paragraph{CodeRepair (program repair).}
The sanitized split of MBPP \citep{austin2021mbpp}. An action is a Python
function; the runtime executes it against the problem's own assertions in a
subprocess and returns the interpreter's output. Failing programs are
single-edit mutations of the reference solution (flip a comparison, shift an
index, drop a \texttt{return}, rename one occurrence of a variable, drop a
negation, and others); a mutant is kept only if the reference passes its tests
and the mutant fails them. Nothing about the error text is authored by us. This
environment also exercises a regime ToolShed cannot: actions are whole functions
rather than one-line calls, and a real traceback \emph{quotes the offending source line back at the model}, a surface-form
echo that appears in agent transcripts without anyone deciding to put it there.

\begin{table}[t]
\centering
\small
\setlength{\tabcolsep}{3.5pt}
\begin{tabular}{l r r r r r r}
\toprule
Environment & items & tasks & ops. & err. & \makecell{action\\(chars)} & \makecell{obs.\\(chars)} \\
\midrule
ToolShed & 275 & 58 & 8 & 12 & 57 & 75 \\
CodeRepair & 266 & 132 & 10 & 4 & 198 & 150 \\
\bottomrule
\end{tabular}
\caption{Probe item pools. \emph{ops.} is the number of perturbation operators used to build failing actions, \emph{err.} the number of distinct error families they provoke. Experiments draw a balanced subsample from these pools; the drawn size is given per study.}
\label{tab:envs}
\end{table}

\subsection{Models}

\NumModels\ instruction-tuned checkpoints spanning \SmallestModel--\LargestModel\
parameters across \NumFamilies\ families (\FamilyList)
\citep{allal2025smollm2,qwen2024qwen25,qwen2025qwen3,grattafiori2024llama3}.
Within-family pairs are included deliberately so that a size trend is
identifiable separately from family effects. All models run in float32 on CPU;
the environment is described in Section~\ref{sec:efficiency}.

\subsection{Protocol}

Probe scoring is teacher-forced and deterministic. For each item we score the
failing action and the reference action under every condition, plus two
distractor actions under the before/after pair. Items are drawn by a
deterministic stratified subsample that balances perturbation operators and
caps items per task, since the number of independent tasks, not the number of items, is what the
confidence intervals depend on.

\paragraph{Statistics.}
Probe items from the same task share a goal, a system prompt and a prefix, so
they are not independent. Every interval we report is a cluster bootstrap
($10{,}000$ resamples) that resamples \emph{tasks} with replacement and carries
all of a task's items along with it. We report the mean difference, its 95\%
interval, Cohen's $d_z$ for the paired differences, and a bootstrap $p$-value
against zero; the family of headline $G$ tests is Holm-corrected across models.
Bold cells in the tables mark intervals excluding zero.

\paragraph{Reproducibility.}
All randomness derives from a single base seed (20260808) through named
sub-streams, so re-running any part of the study reproduces the same items even
if the surrounding loop changed. Raw per-score records, the exact item ids used
by each run and model metadata are written alongside every result file, as is
the software environment for every run made after we started recording it
(Appendix~\ref{app:repro}).

\section{Execution feedback inverts}
\label{sec:results}

\subsection{The headline measurement}

Table~\ref{tab:probe-main-toolshed} reports the corrective gain and its
decomposition on ToolShed. The gain is negative for every model we tested:
recording a failed call and its error message in the transcript makes the model
\emph{more} likely to write that exact call again, by \MeanG\ nats on average.
The intervals exclude zero for \NumModelsNegativeG\ of
\NumModelsToolshed\ models.


\begin{table*}[t]
\centering
\tblfont
\setlength{\tabcolsep}{\tblsep}
\begin{tabular}{l r c c c c c}
\toprule
Model & Params & $G$ & $G$/token & copy & sem & $G_{\text{abs}}$ \\
\midrule
SmolLM2-135M & 0.14B & \textbf{\textminus 23.92\,\ci{\textminus 25.86}{\textminus 22.09}} & \textbf{\textminus 1.21\,\ci{\textminus 1.36}{\textminus 1.06}} & \textbf{21.69\,\ci{19.54}{23.94}} & \textbf{2.23\,\ci{1.59}{2.90}} & \textbf{\textminus 2.24\,\ci{\textminus 3.41}{\textminus 1.05}} \\
SmolLM2-360M & 0.36B & \textbf{\textminus 18.62\,\ci{\textminus 21.00}{\textminus 16.38}} & \textbf{\textminus 0.93\,\ci{\textminus 1.06}{\textminus 0.79}} & \textbf{16.08\,\ci{13.89}{18.43}} & \textbf{2.54\,\ci{1.98}{3.16}} & \textbf{\textminus 3.23\,\ci{\textminus 4.24}{\textminus 2.24}} \\
Qwen2.5-0.5B & 0.49B & \textbf{\textminus 13.91\,\ci{\textminus 16.12}{\textminus 11.82}} & \textbf{\textminus 0.94\,\ci{\textminus 1.14}{\textminus 0.75}} & \textbf{9.91\,\ci{7.59}{12.25}} & \textbf{4.00\,\ci{2.82}{5.28}} & \textbf{\textminus 2.86\,\ci{\textminus 3.71}{\textminus 1.99}} \\
Qwen3-0.6B & 0.60B & \textbf{\textminus 21.50\,\ci{\textminus 25.07}{\textminus 18.18}} & \textbf{\textminus 1.47\,\ci{\textminus 1.74}{\textminus 1.21}} & \textbf{18.25\,\ci{14.75}{21.95}} & \textbf{2.78\,\ci{1.56}{3.96}} & \textbf{\textminus 6.60\,\ci{\textminus 8.07}{\textminus 5.25}} \\
Llama-3.2-1B & 1.24B & \textbf{\textminus 12.41\,\ci{\textminus 15.71}{\textminus 9.62}} & \textbf{\textminus 0.92\,\ci{\textminus 1.21}{\textminus 0.66}} & \textbf{9.42\,\ci{7.23}{11.91}} & \textbf{3.26\,\ci{2.06}{4.62}} & \textbf{\textminus 5.46\,\ci{\textminus 7.49}{\textminus 3.65}} \\
SmolLM2-1.7B & 1.71B & \textbf{\textminus 13.91\,\ci{\textminus 15.53}{\textminus 12.35}} & \textbf{\textminus 0.71\,\ci{\textminus 0.81}{\textminus 0.61}} & \textbf{11.67\,\ci{10.37}{12.97}} & \textbf{2.24\,\ci{1.44}{3.12}} & \textbf{\textminus 4.36\,\ci{\textminus 5.08}{\textminus 3.57}} \\
\bottomrule
\end{tabular}
\caption{Probe results on ToolShed (tool calls). All quantities are in nats and are differences in the summed log-probability of the \emph{same} action string under two contexts. $G>0$ means the harness's failure record made the failed call less likely; $G<0$ is feedback inversion. Cells give the mean over items with a 95\% cluster bootstrap interval (clustered on task); bold marks intervals excluding zero.}
\label{tab:probe-main-toolshed}
\end{table*}

Nats are hard to feel, so Table~\ref{tab:probe-repeat-toolshed} restates the
same measurement two ways. The normalised probability of re-emitting the failed
call, over a fixed four-action candidate set, moves from near the floor before
the attempt to a large fraction after it. And greedy decoding, which is what the agent would actually have written,
reproduces the failed call \emph{token for
token} on \GreedyPostMean\ of items after the failure, against \GreedyPreMean\
before it.


\begin{table*}[t]
\centering
\tblfont
\setlength{\tabcolsep}{\tblsep}
\begin{tabular}{l r c c c c c c}
\toprule
Model & Params & $p_{\text{rep}}$ before & $p_{\text{rep}}$ after & $\Delta p_{\text{rep}}$ & greedy before & greedy after & greedy abs. \\
\midrule
SmolLM2-135M & 0.14B & \textbf{0.142\,\ci{0.086}{0.201}} & \textbf{0.875\,\ci{0.840}{0.910}} & \textbf{0.733\,\ci{0.677}{0.789}} & 0.000\,\ci{0.000}{0.000} & \textbf{0.430\,\ci{0.340}{0.519}} & 0.000\,\ci{0.000}{0.000} \\
SmolLM2-360M & 0.36B & \textbf{0.117\,\ci{0.067}{0.170}} & \textbf{0.753\,\ci{0.684}{0.820}} & \textbf{0.637\,\ci{0.568}{0.708}} & 0.000\,\ci{0.000}{0.000} & \textbf{0.410\,\ci{0.306}{0.521}} & 0.020\,\ci{0.000}{0.050} \\
Qwen2.5-0.5B & 0.49B & \textbf{0.013\,\ci{0.004}{0.028}} & \textbf{0.602\,\ci{0.528}{0.681}} & \textbf{0.589\,\ci{0.516}{0.666}} & 0.000\,\ci{0.000}{0.000} & 0.030\,\ci{0.000}{0.067} & 0.000\,\ci{0.000}{0.000} \\
Qwen3-0.6B & 0.60B & \textbf{0.006\,\ci{0.000}{0.014}} & \textbf{0.296\,\ci{0.196}{0.399}} & \textbf{0.290\,\ci{0.193}{0.393}} & 0.000\,\ci{0.000}{0.000} & 0.000\,\ci{0.000}{0.000} & 0.000\,\ci{0.000}{0.000} \\
Llama-3.2-1B & 1.24B & \textbf{0.009\,\ci{0.000}{0.025}} & \textbf{0.246\,\ci{0.136}{0.363}} & \textbf{0.237\,\ci{0.131}{0.348}} & 0.000\,\ci{0.000}{0.000} & 0.000\,\ci{0.000}{0.000} & 0.000\,\ci{0.000}{0.000} \\
SmolLM2-1.7B & 1.71B & \textbf{0.050\,\ci{0.021}{0.085}} & \textbf{0.476\,\ci{0.393}{0.570}} & \textbf{0.426\,\ci{0.353}{0.506}} & 0.000\,\ci{0.000}{0.000} & \textbf{0.270\,\ci{0.174}{0.381}} & 0.000\,\ci{0.000}{0.000} \\
\bottomrule
\end{tabular}
\caption{Repetition probabilities on toolshed. $p_{\text{rep}}$ is the normalised probability of the failed call among the scored candidate actions; \emph{greedy} is the fraction of items on which greedy decoding reproduces the failed call exactly.}
\label{tab:probe-repeat-toolshed}
\end{table*}

\subsection{It is the string, not the message}

\begin{figure}[t]
\centering
\includegraphics[width=\linewidth]{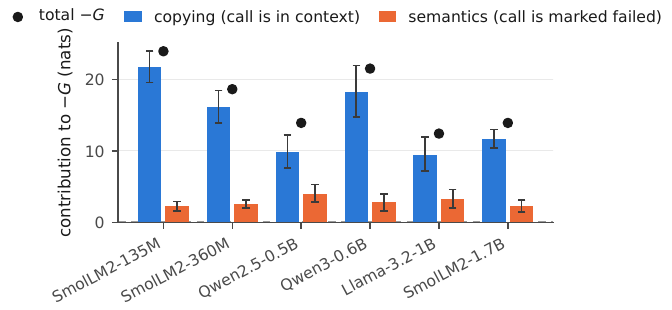}
\caption{Where the inversion comes from. Bars decompose $-\Gain$ into the effect
of the failed call being in the context (blue) and the additional effect of
marking it as failed (orange); black dots are the total. Error bars are 95\%
cluster bootstrap intervals. If the problem were that the model does not
understand the error, the orange bars would be large and negative.}
\label{fig:decomposition}
\end{figure}

Figure~\ref{fig:decomposition} splits $-\Gain$ into the two terms of
Eq.~\eqref{eq:decomp}. The surface-form term $\copyterm$, which is the effect of the call being
present in the transcript at all under a feedback-neutral observation,
accounts for the great majority of the total.

The semantic term $\semterm$ is small, and its sign is not even consistent
across environments. On ToolShed it is slightly positive: the error message
nudges the failed call \emph{further up}, plausibly because explaining what went
wrong requires naming the tool and the offending argument again. On CodeRepair
it is slightly negative, so a Python traceback does push the model away from the
program that produced it. We want to be careful here, because the second case is
the one that matters for the framing: the claim is not that these models cannot
read error messages. Sometimes they can. The claim is that whatever reading
happens is swamped. On CodeRepair the median ratio between the two terms is
about $\CopySemRatioCode{:}1$ in favour of the surface form, against
$\CopySemRatioTool{:}1$ on tool calls.

The robustness table (Table~\ref{tab:probe-robustness}) shows the decomposition
is insensitive to which reference observation defines ``neutral''. Referencing
$\copyterm$ and $\semterm$ to the counterfactual \emph{success} observation
instead of the valence-free one leaves the picture unchanged, and the two
bracket any reasonable notion of neutrality.


\begin{table*}[t]
\centering
\tblfont
\setlength{\tabcolsep}{\tblsep}
\begin{tabular}{l r c c c c c}
\toprule
Model & Params & frac. $G<0$ & $\Delta$margin & repair & copy$^{\checkmark}$ & polarity \\
\midrule
SmolLM2-135M & 0.14B & \textbf{1.00\,\ci{1.00}{1.00}} & \textbf{\textminus 12.58\,\ci{\textminus 14.34}{\textminus 10.90}} & \textbf{11.34\,\ci{10.21}{12.40}} & \textbf{24.46\,\ci{22.57}{26.51}} & \textbf{\textminus 0.54\,\ci{\textminus 0.93}{\textminus 0.15}} \\
SmolLM2-360M & 0.36B & \textbf{1.00\,\ci{1.00}{1.00}} & \textbf{\textminus 12.23\,\ci{\textminus 14.11}{\textminus 10.51}} & \textbf{6.39\,\ci{4.86}{8.02}} & \textbf{17.80\,\ci{15.59}{20.22}} & \textbf{0.82\,\ci{0.20}{1.49}} \\
Qwen2.5-0.5B & 0.49B & \textbf{0.90\,\ci{0.84}{0.95}} & \textbf{\textminus 15.16\,\ci{\textminus 17.71}{\textminus 12.79}} & \textminus 1.25\,\ci{\textminus 2.46}{0.03} & \textbf{11.01\,\ci{8.77}{13.37}} & \textbf{2.90\,\ci{1.73}{4.08}} \\
Qwen3-0.6B & 0.60B & \textbf{0.94\,\ci{0.88}{0.99}} & \textbf{\textminus 17.54\,\ci{\textminus 20.69}{\textminus 14.61}} & \textbf{3.96\,\ci{2.25}{5.75}} & \textbf{21.03\,\ci{17.85}{24.38}} & 0.47\,\ci{\textminus 0.41}{1.28} \\
Llama-3.2-1B & 1.24B & \textbf{1.00\,\ci{1.00}{1.00}} & \textbf{\textminus 9.60\,\ci{\textminus 12.31}{\textminus 7.36}} & \textbf{3.08\,\ci{1.20}{4.96}} & \textbf{12.51\,\ci{9.39}{15.93}} & 0.17\,\ci{\textminus 1.07}{1.30} \\
SmolLM2-1.7B & 1.71B & \textbf{1.00\,\ci{1.00}{1.00}} & \textbf{\textminus 11.61\,\ci{\textminus 13.39}{\textminus 9.97}} & \textbf{2.30\,\ci{1.09}{3.59}} & \textbf{12.92\,\ci{11.48}{14.48}} & \textbf{0.99\,\ci{0.59}{1.41}} \\
\bottomrule
\end{tabular}
\caption{Robustness and secondary quantities. \emph{frac.\ $G<0$} is the assumption-light item-level sign statistic; $\Delta$margin is the change in the gold-minus-failed log-probability gap; \emph{repair} is the change in the correct action's log-probability. copy$^{\checkmark}$ repeats the surface-form term against the counterfactual \emph{success} observation instead of the valence-free one, in case the latter reads as mildly positive; the matching semantic term referenced the same way is algebraically the polarity contrast, so it appears once, in the last column.}
\label{tab:probe-robustness}
\end{table*}

This is the paper's central claim, and it decides the design question posed in
Section~\ref{sec:problem}. The harness is not failing to communicate. It is
supplying, in the same breath, a strong reason to repeat the action and a weak
reason not to.

\subsection{Copying does not simply raise everything}

An obvious objection is that copying should raise the probability of \emph{any}
string that resembles the context, so a negative $\Gain$ may say nothing about
decisions. It does not survive contact with the data. The correct action shares most of its tokens with the failed one, typically
differing in one argument name or one value, so its probability rises too, by \MeanRepair\ nats on
average. What matters for the agent's next move is the \emph{margin} between
them, and the margin moves the wrong way: the gold-minus-failed log-probability
gap shrinks by \MeanMarginShift\ nats once the failure is recorded
(Table~\ref{tab:probe-robustness}, $\Delta$margin). Copying is not neutral
noise; it advantages the wrong action specifically, because the wrong action is
the one that was written down.

\subsection{Scale}

\begin{figure}[t]
\centering
\includegraphics[width=\linewidth]{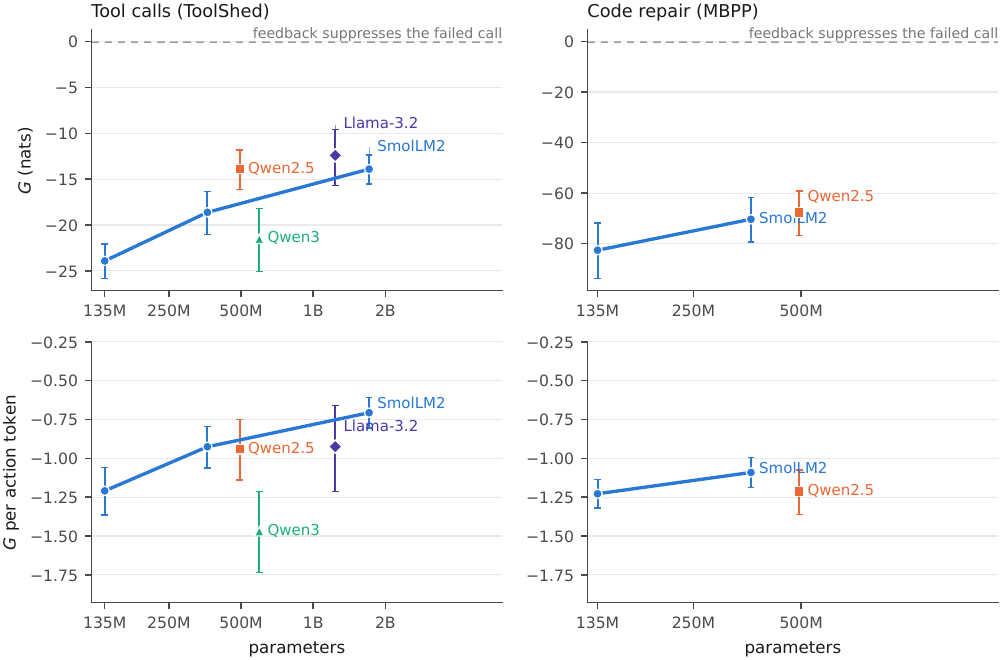}
\caption{Corrective gain against parameter count. Positive values would mean the
failure record does what it is intended to do. \textbf{Top:} raw nats, with an independent axis per environment, since an MBPP program is roughly
five times as many tokens as a tool call and a shared axis would flatten the
left panel.
\textbf{Bottom:} the same quantity per action token, where the axis \emph{is}
shared, because the two environments landing in the same band is the point.
Error bars are 95\% cluster bootstrap intervals; the dashed line is zero.}
\label{fig:scaling}
\end{figure}

Figure~\ref{fig:scaling} plots $\Gain$ against parameter count. Two things are
worth separating here, because they carry different amounts of evidence.

Within a family, where training data is held roughly fixed, the larger
checkpoint has the milder gain, and it does so in both environments
independently. Across families the ordering is not strict: two checkpoints of
similar size but different provenance can differ by several nats, which is what
one should expect when the comparison varies training data as well as parameter
count. We rest the scaling statement on the within-family comparison and treat
the pooled fit as descriptive.

What the decomposition adds is that the weakening is driven by $\copyterm$
rather than by $\semterm$. Larger models copy less; they do not read the error
better.

A log-linear fit over the range we can run has slope \ScalingSlope\ nats per
decade of parameters ($R^2 = \ScalingRsq$) and reaches zero at
\ZeroCrossing. We report that number in order to argue against it. Refitting
with each model held out in turn moves the crossing from
\ZeroCrossingLoMin\ to \ZeroCrossingLoMax, a factor of
\ZeroCrossingLoRatio\ decided by which six checkpoints happened to be
affordable. That is what an extrapolation an order of magnitude beyond the
measured range is worth here, and it is why we treat the pooled fit as
descriptive of a direction rather than as a prediction of a threshold.

Two things could be true at 8B or 70B that our data cannot distinguish: the
copying term could keep shrinking on the same line, or the semantic term could
begin to dominate. Since our measurement costs a few hundred forward passes per
model, this is a cheap question for anyone with the hardware to answer, and we
would rather flag it than guess at it.

\subsection{The second environment, and a unit that travels}

The same pattern appears in program repair, where the observations are the
Python interpreter's own output rather than anything we wrote
(Table~\ref{tab:probe-main-coderepair}). Actions here are whole functions, an
order of magnitude longer than tool calls, and real tracebacks quote the
offending source line, so the surface form is re-injected without anyone
deciding to do so. The gain is \MeanGCode\ nats on average across
\NumModelsCode\ models.

Raw nats are not comparable between the two environments, since a program is
roughly five times as many tokens as a tool call. Divided by action length they
are, and the result is worth stating: the corrective gain is close to
\emph{one nat per action token} in both environments. A single quantity of that
size, arriving at the same value from a twelve-tool simulated workspace and from
MBPP with real tracebacks, is the best evidence we have that the effect is a
property of how transcripts are assembled rather than of anything specific to
either setting.

Two other quantities in Table~\ref{tab:probe-robustness} deserve mention because
they need no distributional assumptions at all. The fraction of \emph{individual
items} on which the gain is negative is at or near one for every model. And
under the abstracted harness on CodeRepair the gain is positive: removing the verbatim program while keeping the
diagnosis does not merely reduce the inversion, it reverses it.


\begin{table*}[t]
\centering
\tblfont
\setlength{\tabcolsep}{\tblsep}
\begin{tabular}{l r c c c c c}
\toprule
Model & Params & $G$ & $G$/token & copy & sem & $G_{\text{abs}}$ \\
\midrule
SmolLM2-135M & 0.14B & \textbf{\textminus 82.61\,\ci{\textminus 93.89}{\textminus 71.79}} & \textbf{\textminus 1.23\,\ci{\textminus 1.32}{\textminus 1.14}} & \textbf{84.00\,\ci{73.16}{95.33}} & \textbf{\textminus 1.39\,\ci{\textminus 1.88}{\textminus 0.91}} & \textbf{1.56\,\ci{0.93}{2.23}} \\
SmolLM2-360M & 0.36B & \textbf{\textminus 70.35\,\ci{\textminus 79.40}{\textminus 61.71}} & \textbf{\textminus 1.09\,\ci{\textminus 1.19}{\textminus 1.00}} & \textbf{71.02\,\ci{62.08}{80.28}} & \textbf{\textminus 0.67\,\ci{\textminus 1.16}{\textminus 0.19}} & \textbf{\textminus 0.80\,\ci{\textminus 1.23}{\textminus 0.40}} \\
Qwen2.5-0.5B & 0.49B & \textbf{\textminus 67.68\,\ci{\textminus 76.71}{\textminus 59.15}} & \textbf{\textminus 1.21\,\ci{\textminus 1.36}{\textminus 1.08}} & \textbf{69.50\,\ci{61.01}{78.50}} & \textbf{\textminus 1.82\,\ci{\textminus 2.48}{\textminus 1.19}} & \textbf{\textminus 0.51\,\ci{\textminus 0.95}{\textminus 0.08}} \\
\bottomrule
\end{tabular}
\caption{Probe results on CodeRepair (MBPP program repair). All quantities are in nats and are differences in the summed log-probability of the \emph{same} action string under two contexts. $G>0$ means the harness's failure record made the failed call less likely; $G<0$ is feedback inversion. Cells give the mean over items with a 95\% cluster bootstrap interval (clustered on task); bold marks intervals excluding zero.}
\label{tab:probe-main-coderepair}
\end{table*}

\section{What moves the effect, and what does not}
\label{sec:analysis}

\begin{figure}[t]
\centering
\includegraphics[width=\linewidth]{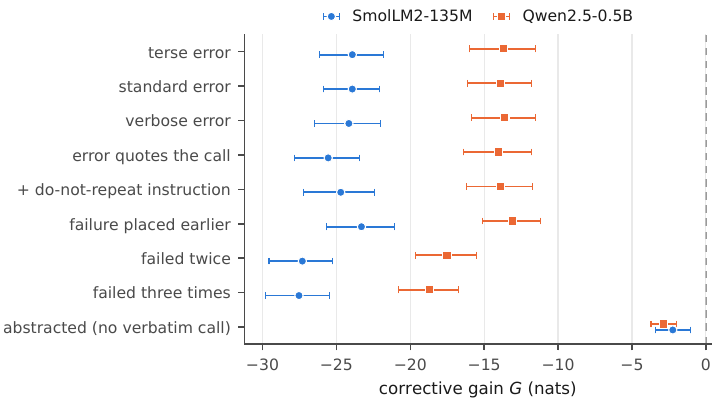}
\caption{Corrective gain under harness variants. Points right of the dashed line
would mean the failure record suppresses the failed call. Manipulations of the
message (top four rows) move the estimate very little; the manipulation that
removes the call's surface form (bottom row) moves it by an order of magnitude.}
\label{fig:manipulations}
\end{figure}

Section~\ref{sec:results} attributes the inversion to the failed call's surface
form. If that attribution is right, the effect should respond to manipulations
of the \emph{form} and be largely indifferent to manipulations of the
\emph{message}. Table~\ref{tab:probe-manipulations} and
Figure~\ref{fig:manipulations} test exactly that. Every row uses the same items
and the same failing call; only how the failure is written into the transcript
changes.

Throughout this section we report each variant as a \emph{paired} change in
$\log \pi(\afail)$ relative to the standard harness, computed item by item.
Comparing two independently estimated gains would confound the manipulation with
which items each variant happened to be scored on; pairing removes that, and it
also removes item difficulty, which is the dominant source of variance here.
Positive means the variant makes repetition \emph{more} likely.


\begin{table*}[t]
\centering
\tblfont
\setlength{\tabcolsep}{\tblsep}
\begin{tabular}{l c c}
\toprule
Harness variant & SmolLM2-135M & Qwen2.5-0.5B \\
\midrule
terse error text & \textminus 0.32\,\ci{\textminus 0.72}{0.02} & \textminus 0.23\,\ci{\textminus 0.64}{0.20} \\
verbose error text & \textminus 0.09\,\ci{\textminus 0.32}{0.14} & \textminus 0.31\,\ci{\textminus 0.65}{0.02} \\
error quotes the failed call & \textbf{1.31\,\ci{1.12}{1.51}} & 0.10\,\ci{\textminus 0.22}{0.44} \\
failure placed before the successful steps & \textbf{\textminus 0.94\,\ci{\textminus 1.45}{\textminus 0.48}} & \textbf{\textminus 0.82\,\ci{\textminus 1.36}{\textminus 0.26}} \\
+ ``do not repeat'' instruction & \textbf{0.16\,\ci{0.06}{0.27}} & \textminus 0.02\,\ci{\textminus 0.20}{0.15} \\
the same call failed twice & \textbf{3.06\,\ci{2.78}{3.36}} & \textbf{3.61\,\ci{3.12}{4.08}} \\
the same call failed three times & \textbf{3.29\,\ci{3.01}{3.59}} & \textbf{4.78\,\ci{4.18}{5.37}} \\
placebo: a \emph{different} call failed & \textbf{\textminus 15.71\,\ci{\textminus 17.38}{\textminus 14.23}} & \textbf{\textminus 11.71\,\ci{\textminus 13.28}{\textminus 10.18}} \\
abstracted failure (no verbatim call) & \textbf{\textminus 21.68\,\ci{\textminus 23.20}{\textminus 20.14}} & \textbf{\textminus 11.05\,\ci{\textminus 12.80}{\textminus 9.39}} \\
\quad bare marker, no diagnosis & \textbf{\textminus 24.23\,\ci{\textminus 26.30}{\textminus 22.32}} & \textbf{\textminus 13.01\,\ci{\textminus 15.04}{\textminus 11.10}} \\
\bottomrule
\end{tabular}
\caption{Harness variants, as \emph{paired} changes in the log-probability of re-emitting the failed call, relative to the standard harness on the same items. Positive means the variant makes repetition \emph{more} likely; negative means less. Every row uses the same items and the same failed call, and only how the failure is written into the transcript changes.}
\label{tab:probe-manipulations}
\end{table*}

\paragraph{Message content is nearly inert.}
Shortening the error to a bare category name changes the log-probability of
repeating by \DTerseCI\ nats; expanding it with the tool signature and a
corrective hint changes it by \DVerboseCI. Both intervals contain zero. The
verbose rendering explicitly names the argument the model should have used, which is about as informative
as an error message can be without solving the task, and it does not
measurably move the quantity that decides whether the call gets
written again.

\paragraph{Occurrence count is not inert.}
When the runtime echoes the offending call inside the error message, as a
Python traceback does by default and as several agent frameworks do by
convention, the failed call appears twice instead of once, and repetition
becomes \DEchoCI\ nats more likely. The interval excludes zero. Set against
the near-zero effect of rewriting the message, this is the sharpest contrast we
have between form and content: adding information changes nothing, adding a
second copy of the string changes something.

It is also a small, actionable finding on its own. A runtime that helpfully
quotes back what you tried is, for a small model, making the situation slightly
worse.

\paragraph{Repetition compounds.}
When the same call fails twice the effect grows by \DTwiceCI\ nats, and three
times by \DThriceCI. That is the situation an agent is already in by the time
anyone notices it is stuck, and each additional failure digs the hole deeper.
This is the self-reinforcement of \citet{xu2022breaktheloop} operating inside an
agent trajectory, with the difference that here an explicit corrective signal is
present at every repetition and does not arrest it.

\paragraph{Recency is not the explanation.}
Moving the failed step to \emph{before} the successful prefix steps, so that it
is no longer the most recent thing in the transcript, helps by \DEarlyCI\ nats, a real effect and about three percent of the
total. Whatever is happening
is not simply the last turn dominating.

\paragraph{Telling the model not to repeat does not help.}
Adding a system-prompt instruction not to repeat a call that has already failed
changes the log-probability of repeating by \DInstructionCI\ nats. The
instruction is unambiguous, it sits in the same context window, and the model is
instruction-tuned. The effect is small but its sign is the wrong one, which is
what the negative-instruction literature would predict
\citep{castricato2024pinkelephants,negation2025pinkelephant}: a prohibition that
has to name a string cannot easily beat the presence of that string.
Section~\ref{sec:agent} reports the same intervention in free-running rollouts,
where it also fails to reduce repetition, but where it does change behaviour
in a way this measurement does not anticipate.

\paragraph{Removing the surface form does.}
Replacing the verbatim call with a runtime-generated description of the
failure, which keeps the diagnosis and drops the token sequence, moves the
log-probability of repeating
by \DAbstractCI\ nats, which is very nearly the whole effect, and drives the
exact greedy repeat rate to zero for every model
(Table~\ref{tab:probe-repeat-toolshed}). The description is produced
deterministically from the runtime's own error metadata; no additional model
call is involved, so this is a change to the harness, not to the agent.

\subsection{Controls}

\paragraph{A placebo: someone else's failure.}
The strongest version of the ``you have just rediscovered copying'' objection is
that any failure in the context might raise the probability of any plausible
action. We test it directly. In the \emph{fail\_other} condition the transcript
records a \emph{different} plausible-but-wrong call as having failed, with its
own genuine error message; the action we score was never written.

Relative to the standard harness, the placebo recovers \DPlaceboCI\ nats, a large majority of the effect,
but not all of it. The reason it does not
recover all of it is worth being explicit about: the distractor is another
perturbation of the
\emph{same} reference call, so it shares most of its tokens with the action being
scored, and the placebo therefore leaks copying by construction. That makes it a
conservative control rather than a broken one. It bounds the string-specific
component from below at \StringSpecificBound\ nats, and we report it that way
rather than attributing all of $\Gain$ to the exact string. A placebo drawn from
an unrelated call would tighten the bound and is the obvious refinement.

\paragraph{Does the abstraction work because of what it says?}
Our abstraction replaces the call with a short diagnosis, and we wrote those glosses, so perhaps it works because the glosses are good,
which would not generalise. \emph{abstract\_min} strips them: the transcript gets
\texttt{[attempt 1 failed]} and nothing else, no call and no diagnosis. It
recovers \DAbstractMinCI\ nats, against \DAbstractMatchedCI\ for the full abstraction on the same models, if
anything slightly more. The benefit is
the absence of the string, not our wording, which is what the decomposition
predicts and the version of the result that transfers to other runtimes.

That does not make the diagnosis worthless, since it is what the model needs
in order to write a \emph{different} call rather than merely a different
string, but it is not what is doing the work here.

\paragraph{Is the ``neutral'' observation really neutral?}
``Call recorded.'' might read as mildly positive, which would make $\copyterm$
absorb some of what belongs to $\semterm$. Table~\ref{tab:probe-robustness}
therefore repeats the decomposition using the counterfactual \emph{success}
observation as the reference instead. The two references bracket any reasonable
notion of neutrality, and the conclusion should not depend on which is chosen.

\paragraph{Length.}
The failure observation is longer than the valence-free one, so part of
$\semterm$ could be a position effect rather than a content effect. Repeating
the decomposition against a neutral observation padded to the same token length
with valence-free trace identifiers separates the two, and the answer is clean
in both directions. The surface-form term is unmoved, at \MeanCopy\ against a short neutral and
\CopyPad\ against a length-matched one, so the effect the paper rests on is
not an artefact of context length. The semantic term shrinks from
\MeanSem\ to \SemPad, meaning most of the small positive value it had was
indeed position rather than content. What survives is a corrective term of under
a nat, against a surface-form term of more than twenty.

\begin{figure}[t]
\centering
\includegraphics[width=\linewidth]{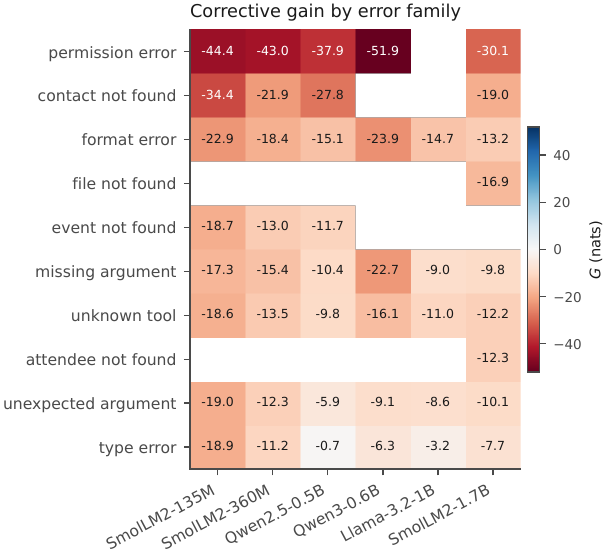}
\caption{Corrective gain by error family (ToolShed), one cell per
model\,$\times$\,family. The scale is diverging about zero, so red is feedback
inversion and blue would be a harness behaving as intended; no cell is blue.
Blank cells are model/family pairs the schedule did not reach.}
\label{fig:error-types}
\end{figure}

\paragraph{Error family.}
Figure~\ref{fig:error-types} and Table~\ref{tab:by-error} break the gain down by
the kind of error the call provoked. The effect is present in every family; a
result concentrated in one would have suggested that the wording of one error
message, rather than the mechanism we describe, was responsible.

\subsection{Does copying just raise everything?}

We return to this because it is the objection a careful reader should raise
twice. Copying from context is indiscriminate, so the correct action, which usually differs from the failed one by a single
argument name or value, should also become more likely, and it does. Three observations keep the result from
collapsing into that.

First, the quantity that governs the agent's next move is the difference, and
the difference moves against the correct action
(Table~\ref{tab:probe-robustness}, $\Delta$margin). Second, the normalised
repeat probability is computed over a fixed candidate set that includes the
correct action and two alternative wrong actions, so a uniform lift cancels out;
it still rises sharply. Third, the greedy readout is a statement about the
argmax, which is immune to uniform lifts by construction: after the failure, the
single most likely continuation \emph{is} the failed call on a large fraction of
items, and before it, on none.

\section{Does it matter when the agent drives?}
\label{sec:agent}

The probe holds the trajectory fixed and measures a distribution. The obvious
next question is whether any of it survives when the model chooses its own
actions, makes its own mistakes, and has several steps to recover from them.

\subsection{Setup}

Models act freely in ToolShed over \AgentNumTasks\ tasks, for up to
\AgentMaxSteps\ steps each, with greedy decoding. The task mix interleaves
single-call tasks (resolve a contact, write a file, set a reminder) with the two-
and three-call tasks used in the probe; without the easy tasks the smallest
models score zero under every harness and the comparison would be made on a
floor. A one-line format demonstration is appended to the system prompt,
identically in every condition. A task counts as solved only if the environment's goal predicate holds. There
is no partial credit and no model judges anything.

Two details of that demonstration are worth stating, because getting them wrong
cost us a full run. It shows the reply alone, with no speaker labels: an earlier
version presented a two-turn dialogue, and models copied the ``you:'' label into
their own replies, which the parser then rejected. The rate at which they did so
varied by harness, so a presentational choice had become a confound with the
comparison. Relatedly, the parser strips a leading speaker or action label
before parsing. That is the right behaviour independent of the bug: a harness
should not score a chat model's formatting habit as a tool-use failure. The
superseded rollouts are retained in the repository with an explanation rather
than deleted.

We compare \AgentNumHarnesses\ harnesses, which differ only in what the model
sees and in how its next token is chosen:
\emph{verbatim} (the standard transcript),
\emph{verbatim+instruction},
\emph{drop} (the failed step deleted, i.e.\ clean restart),
\emph{abstract} (failed calls replaced by a runtime-generated description),
\emph{verbatim+ban} (transcript unchanged, previously-failed strings blocked at
the decoder), and
\emph{abstract+ban}.

\subsection{Metrics}

Task success is the headline, but it is a coarse instrument at these model
sizes, so we report the mechanism alongside it.

\emph{Ended in a loop} is the fraction of rollouts that exhaust the step budget
having repeated an action. It is the outcome practitioners actually complain
about.

\emph{Exact repeat rate} is the fraction of failed actions that reproduce, byte
for byte, an action that already failed in the same rollout.

\emph{Canonical repeat rate} applies the same count after parsing the call and
sorting its keyword arguments, so a model that evades the ban by reordering
arguments or changing whitespace is still counted as repeating. That metric
exists to keep the ban honest: a decoder constraint trivially drives the exact
rate to zero, and the question is whether the model then does something useful
or merely paraphrases its mistake.


\begin{table*}[t]
\centering
\small
\setlength{\tabcolsep}{4pt}
\begin{tabular}{l c c c}
\toprule
 & \multicolumn{3}{c}{Qwen2.5-0.5B} \\
Harness & success & exact rep. & canon. rep. \\
\midrule
verbatim (standard) & 0.42\,\ci{0.21}{0.62} & 0.31\,\ci{0.15}{0.47} & 0.31\,\ci{0.15}{0.47} \\
\quad + ``do not repeat'' & 0.58\,\ci{0.38}{0.79} & 0.24\,\ci{0.08}{0.42} & 0.24\,\ci{0.08}{0.42} \\
abstract & 0.33\,\ci{0.17}{0.54} & 0.16\,\ci{0.04}{0.31} & 0.16\,\ci{0.04}{0.31} \\
drop & 0.33\,\ci{0.17}{0.54} & 0.80\,\ci{0.79}{0.81} & 0.80\,\ci{0.79}{0.81} \\
verbatim + ban & 0.42\,\ci{0.21}{0.62} & 0.08\,\ci{0.00}{0.15} & 0.08\,\ci{0.00}{0.15} \\
abstract + ban & 0.33\,\ci{0.17}{0.54} & 0.07\,\ci{0.00}{0.14} & 0.07\,\ci{0.00}{0.14} \\
\bottomrule
\end{tabular}
\caption{End-to-end rollouts on ToolShed. Success is the environment's goal predicate; \emph{exact rep.} is the fraction of failed actions that repeat an earlier failed action byte for byte, and \emph{canon. rep.} applies the same count after parsing and normalising the call, so paraphrases of a banned string are still counted. Brackets give 95\% cluster bootstrap intervals over tasks.}
\label{tab:agent-main}
\end{table*}

\begin{figure*}[t]
\centering
\includegraphics[width=0.9\linewidth]{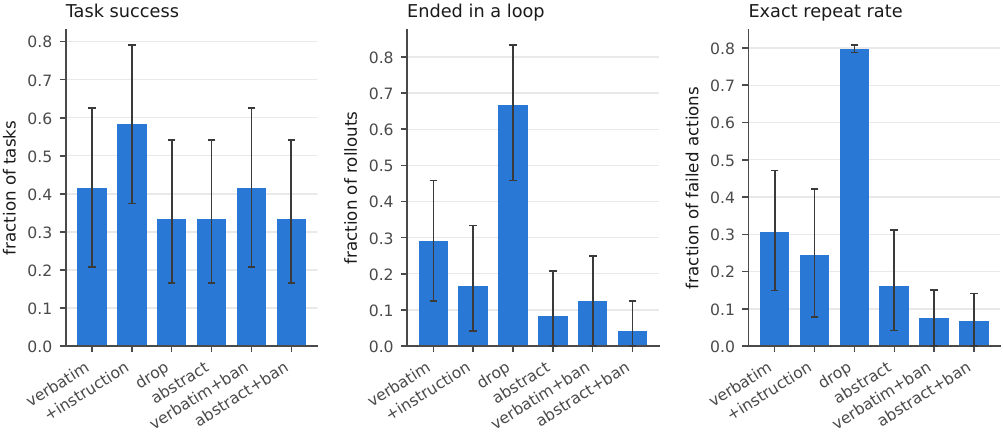}
\caption{End-to-end rollouts, by harness. \textbf{Left:} task success.
\textbf{Middle:} the fraction of rollouts that end in a loop (step limit reached
with a repeated action). \textbf{Right:} the fraction of failed actions that
repeat an earlier failed action byte for byte. The two repetition panels order
the harnesses the same way, and it is the order the decomposition predicts:
interventions acting on the failed call's surface form at the good end, the
harness that deletes the failure outright at the bad end, the standard harness
in between. Task success does not follow that order, which
Section~\ref{sec:failure} explains. Error bars are 95\% cluster bootstrap
intervals over tasks.}
\label{fig:agent}
\end{figure*}

\subsection{Results}

\paragraph{The baseline behaves as the probe predicts.}
Under the standard harness the agent solves \AgentVerbatimSuccess\ of tasks, and
\AgentVerbatimRepeat\ of its failed actions are byte-for-byte repeats of an
action that already failed in the same rollout. Repetition is not a rare
pathology here; it is a substantial fraction of everything the agent does wrong.

\paragraph{Telling the model not to repeat does not reduce repetition, but it
does something else.}
Adding the prohibition to the system prompt moves the exact repeat rate by
\DeltaInstrRepeat\ points on matched tasks, an interval containing zero. The
probe reached the same verdict from a different direction
(Section~\ref{sec:analysis}): the instruction leaves the log-probability of
re-emitting the failed call essentially where it was, if anything nudging it the
wrong way.

Task success, however, moves by \DeltaInstrSuccess\ points, and that interval
excludes zero. The instruction also makes the agent \emph{cheaper}:
\AgentInstrTokens\ generated tokens per rollout against
\AgentVerbatimTokens, in fewer steps. We report this because it is what we
measured, and we decline to build a mechanism on it: with
\AgentNumTasks\ tasks the effect is real but the explanation is not identified,
and the most plausible readings, that the instruction makes the model more
careful in general or less willing to abandon a task early, are not about
repetition at all. The claim we do make is the narrow one the probe and the rollouts agree
on: a natural-language prohibition is not a repetition remedy.

\paragraph{Clearing the context is not a fix; it is the worst case.}
The remedy recommended by prior work on context contamination is to clear the
failed attempt before retrying \citep{yang2026contamination}. In our taxonomy
that is the \emph{drop} harness, and it is the harness that repeats most. The
exact repeat rate rises from \AgentVerbatimRepeat\ to \AgentDropRepeat, a
paired difference of \DeltaDropRepeat\ points whose interval is well clear of
zero, and task success does not improve (\DeltaDropSuccess).

The mechanism is worth stating plainly, because it reframes what the fix has to
be. Deleting the failed step restores the \emph{exact} context that produced the
failure. A deterministic policy in an identical context emits an identical action, so
clean restart does not remove the problem, it guarantees it. The
agent then spends its whole budget re-deriving the same call
(\AgentDropTokens\ generated tokens per rollout against
\AgentVerbatimTokens\ for the standard harness) and its rate of producing a
valid call falls.

This is specific to deterministic decoding, and we say so rather than
generalising it: with temperature sampling, resampling would supply the
variation the context no longer does. Greedy decoding is nonetheless a common
deployment choice, and the point survives in weaker form for any low-temperature
agent.

The three harnesses now form a clean design. \emph{verbatim} changes the context
using the failed string, which pulls the model back toward it. \emph{drop} does
not change the context at all, which leaves the model exactly where it was.
The requirement is therefore narrower than ``remove the failed action'':

\begin{quote}
The context after a failure must differ from the context before it, and the
difference must not be the failed action itself.
\end{quote}

\paragraph{Both surface-form interventions reduce repetition; the decoder ban
does so most cleanly.}
Blocking previously-failed strings at the decoder takes the exact repeat rate
from \AgentVerbatimRepeat\ to \AgentVerbBanRepeat, a paired difference of
\DeltaVerbBanRepeat\ points, with an interval clear of zero. The canonical rate
falls by the same amount, so the model is not paraphrasing its way around the constraint; it writes
genuinely different calls. Task success is
unchanged (\DeltaVerbBanSuccess), and the intervention costs no generated tokens
(\AgentVerbBanTokens\ against \AgentVerbatimTokens).

Replacing the call with a runtime-generated description does the same thing
through the context rather than the decoder, and lands between the ban and the
baseline: \AgentAbstractRepeat\ exact repeats, a paired difference of
\DeltaAbstractRepeat\ points whose interval touches zero at this sample size.

\paragraph{Neither surface-form intervention improves task success, and we say
so.}
The abstraction moves success by \DeltaAbstractSuccess\ points and the ban by
\DeltaVerbBanSuccess. This is the result we would most like to have come out
otherwise, and it is worth being direct about what it does and does not
undermine.

The interventions were derived from a mechanism. They move that mechanism, by
roughly the predicted amount and in the predicted direction, and the harness
that our decomposition says should be worst is indeed dramatically worst. What they do not do is convert that into task success at this scale, which
Section~\ref{sec:failure} explains: \ProseShare\ of this model's actions are
prose where a call was expected, a failure no arrangement of the transcript
repairs. Removing the loop returns the agent's budget to it; whether it can
spend that budget well is a question about the model, not the harness.

The instruction is the instructive contrast. It raises success
(\DeltaInstrSuccess\ points) while leaving repetition where it was; the
surface-form interventions do the reverse. They act on different things. Only
the second kind is evidence about the mechanism this paper is concerned with,
and a practitioner who wants task success from a 0.5B agent should not expect
either to be the whole answer.

\paragraph{The ban stops the loop but does not solve the task.}
Applying the decoder ban to the unmodified transcript cuts exact repetition to
near nothing (\DeltaVerbBanRepeat\ points, from \AgentVerbatimRepeat\ to
\AgentVerbBanRepeat) and eliminates loops
(\AgentVerbBanStuck\ of rollouts against \AgentVerbatimStuck), at a cost of
\AgentVerbBanTokens\ generated tokens per rollout against
\AgentVerbatimTokens. Task success does not move at all
(\DeltaVerbBanSuccess).

That combination is worth dwelling on, because it separates two things the
paper could otherwise be read as conflating. The canonical repeat rate equals
the exact rate in every harness we ran, to the digit, so the residual
repetition is not the model paraphrasing its way around the constraint. It is
the ban's own boundary: the mask is over token sequences in the raw generation,
while a repeat is counted over the parsed action, and the two come apart when
the generation is cut off by the token cap or when the same visible string is
reached by a different token path. Inspecting the three rollouts (of
\AgentNumTasks) in which any exact repeat survives the ban, the repeated string is in every case one the parser rejected, either
unterminated prose or a call with an unbalanced parenthesis, and never a
well-formed call. Forbidding a string
prevents the agent from wasting its budget, and nothing more. Removing the
string from the context lets the agent reconsider.

\paragraph{The two interventions are complementary, not redundant.}
Applying the ban on top of the abstraction reduces exact repetition further,
from \AgentAbstractRepeat\ to \AgentAbsBanRepeat\ (\DeltaAbsBanRepeat\ points
against the standard harness, an interval clear of zero where the abstraction
alone is not). The instrumentation says why: under the abstracted transcript the
ban still fires, on roughly half as many decoding steps as it does under the
verbatim transcript, but far from never.

That is worth stating because it refines the mechanism. Removing the failed
call from the context removes the model's opportunity to \emph{copy} it, but
not its ability to \emph{re-derive} it: the goal, the tool schemas and the
successful prefix are all still there, and they are what produced the wrong call
in the first place. The context edit and the decoder constraint therefore catch different things,
one the copying and the other the residue, and a harness that wants neither
should do both. Neither costs a generated token.

\paragraph{Summary of the ordering.}
Ranked by the fraction of rollouts that end in a loop: abstract+ban
(\AgentAbsBanStuck), abstract (\AgentAbstractStuck) and verbatim+ban
(\AgentVerbBanStuck) at the good end; the ``do not repeat'' instruction at
\AgentInstrStuck; the standard harness at \AgentVerbatimStuck; and clean
restart worst at \AgentDropStuck. Ranking by exact repeat rate gives the same
order. Interventions that act on the failed call's surface form occupy the good
end, the harness that deletes the failure outright occupies the bad end, and the standard harness sits between. That is the ordering the decomposition
predicts, arrived at without reference to it.

Task success does not follow that ordering, and we would rather flag the
mismatch than bury it. The best harness for success is the one that does least
about repetition. Section~
ef{sec:failure} gives the reconciliation: at this
scale repetition is a large share of what goes wrong but not the largest, and an
intervention that removes it cannot remove the rest.

\subsection{Does the probe agree with the rollouts?}

Within this model the two studies agree on every comparison they both make. The
probe says the abstraction removes almost all of the inversion and drives the
exact greedy repeat rate to zero; the rollouts show the same, and add that task
success rises. The probe says the natural-language prohibition does not help
(\DInstructionCI\ nats, slightly the wrong way); the rollouts show it increasing
exact repetition. The probe's clean-restart baseline is the \emph{pre}
condition by construction, and the rollouts show why restoring that context is
not a remedy.

What we cannot yet do is test whether the probe \emph{predicts across models},
that is, whether a model with a more negative $\Gain$ loops more in practice. That needs
rollouts for several models, and rollouts are three orders of magnitude more
expensive than probe items on the hardware available here (Section~\ref{sec:efficiency}).
It is the experiment we would run first with a GPU, and it is the one that would
turn the probe from an explanation into a cheap diagnostic: a harness designer
could then measure $\Gain$ on a few hundred items rather than running a full
agent evaluation.

\section{Cost}
\label{sec:efficiency}

\paragraph{Hardware.}
Everything in this paper was produced on one laptop CPU
(\CPUName, four physical cores, 17\,GB RAM) with no GPU, using
PyTorch \TorchVersion\ and Transformers \TransformersVersions\ (see
Appendix~\ref{app:repro}), in float32. We
report this not as a hardship but because it constrains what claims we are
entitled to make, and because it makes the artefact reproducible by anyone.

\paragraph{What the interventions cost.}
The two interventions we recommend are close to free, which is the main reason
they are worth reporting at all.

The \emph{abstract} harness \emph{reduces} context length: a runtime-generated
description of a failure is shorter than the failed call plus its error message.
Its only requirement is that the runtime knows what kind of error it produced,
which any harness that formats an error message already does. It costs no extra
model call and adds no latency.

The \emph{ban} adds one pass over the ban list per decoding step and no tokens
at all. In our implementation the ban list holds at most a handful of short
token sequences, and the added time is below the noise floor of our wall-clock
measurements. Table~\ref{tab:cost} reports generated tokens, prompt tokens and
wall-clock per rollout for each harness so that this can be checked rather than
taken on trust; note that a harness which loops less also \emph{finishes
earlier}, so the interventions tend to reduce total cost rather than add to it.


\begin{table}[t]
\centering
\small
\begin{tabular}{l r r r r}
\toprule
 & \multicolumn{4}{c}{Qwen2.5-0.5B} \\
Harness & gen & prompt & ctx/step & steps \\
\midrule
verbatim (standard) & 67 & 2764 & 633 & 4 \\
\quad + ``do not repeat'' & 47 & 2210 & 646 & 3 \\
abstract & 64 & 2602 & 616 & 4 \\
drop & 69 & 2587 & 586 & 4 \\
verbatim + ban & 69 & 2764 & 633 & 4 \\
abstract + ban & 66 & 2602 & 616 & 4 \\
\bottomrule
\end{tabular}
\caption{Cost per rollout. \emph{gen} and \emph{prompt} are tokens summed over a rollout's steps, so a harness that takes more steps accumulates more of both; \emph{ctx/step} divides the prompt total by the number of steps and is the fair comparison of context size. The decoder ban adds no tokens at all. The abstraction produces a slightly shorter context per step while taking more steps, because it does not terminate early on a loop. Wall-clock is omitted: these runs shared a CPU with other jobs, so it measures scheduling rather than the harness.}
\label{tab:cost}
\end{table}

\paragraph{What the study cost.}
The probe is deliberately built out of teacher-forced scoring rather than
generation. A single decoding step on this machine costs roughly as much as
thirty scored positions, because at batch size one a forward pass is dominated
by per-layer dispatch rather than arithmetic; measuring a distribution is
therefore two orders of magnitude cheaper than sampling from it. Prefix reuse
across conditions removes \CacheSaving\ of the token positions a naive
implementation would push through the network, since our contexts share a long
system prompt and differ only in a short tail.

We also record what we could \emph{not} afford. Multi-threaded inference is
unavailable on this machine for a reason unrelated to the science
(Section~\ref{sec:limitations}), so all parallelism is across processes. Models
above roughly 2B parameters do not fit alongside the rest of the workload in
float32, which is the binding constraint on the top of our ladder.

\section{Where the interventions fail}
\label{sec:failure}

A harness change that only ever helps would be suspicious. Three regimes where
ours do not, or where the phenomenon itself weakens, are worth stating plainly.

\paragraph{The ban displaces repetition into paraphrase.}
Blocking a token sequence guarantees the model cannot emit that sequence; it
does not make the model change its mind. In rollouts we observe the predicted
consequence directly: a model whose exact repeat is blocked will sometimes
produce a near-identical call with a reordered argument, different quoting, or a
trivially altered value. This is why we report the canonical repeat rate, which
counts those as repeats, alongside the exact rate. The gap between the two
metrics is the honest measure of how much the ban actually accomplishes, and it
is the reason we do not recommend the ban on its own.

Figure~\ref{fig:qualitative-ban} shows a pair of trajectories from the same
task and model under \emph{verbatim} and \emph{verbatim+ban}, and it is worth
reading closely because it makes the mechanism visible in a way the aggregates
cannot. Under the standard harness the model emits one wrong call and then
repeats it, byte for byte, five times. Under the ban it cannot, and what it
does instead is not to reconsider the call: it keeps the same call and starts
appending fragments of the error message to it as an extra argument
(\texttt{error='write\_file: SyntaxError: could not parse'}). The exact repeat
rate for that rollout is zero, and nothing has been fixed. Whatever is
producing this behaviour is copying from the transcript, and forbidding one
particular string from the transcript redirects it to the nearest other string
rather than turning it off.

\begin{figure*}[t]
\centering
\begin{minipage}{\textwidth}
\textbf{qwen2.5-0.5b, harness=verbatim}
{\verbfont\begin{verbatim}
goal: agent-0011  solved=False
1. [ok ] convert_units(value=88, from_unit='km', to_unit='mi')
2. [ERR] write_file(path='reports/conversion.md', content=str(54.6806))
3. [ERR] write_file(path='reports/conversion.md', content=str(54.6806))
4. [ERR] write_file(path='reports/conversion.md', content=str(54.6806))
5. [ERR] write_file(path='reports/conversion.md', content=str(54.6806))
6. [ERR] write_file(path='reports/conversion.md', content=str(54.6806))
\end{verbatim}}

\textbf{qwen2.5-0.5b, harness=verbatim-ban}
{\verbfont\begin{verbatim}
goal: agent-0011  solved=False
1. [ok ] convert_units(value=88, from_unit='km', to_unit='mi')
2. [ERR] write_file(path='reports/conversion.md', content=str(54.6806))
3. [ERR] write_file(path='reports/conversion.md', content=str(54.6806), error='write_file: argument value ~
4. [ERR] write_file(path='reports/conversion.md', content=str(54.6806), error='write_file: SyntaxError: co~
5. [ERR] write_file(path='reports/conversion.md', content=str(54.6806), error='write_file: SyntaxError: co~
6. [ERR] write_file(path='reports/conversion.md', content=str(54.6806), error='write_file: argument value ~
\end{verbatim}}
\end{minipage}
\caption{The same task and model under the standard harness and under the decoder ban. Selected by rule, not by eye: the task with the most exact repeats under the standard harness for which a matching rollout exists under the comparison harness. Actions are shown as the parser received them.}
\label{fig:qualitative-ban}
\end{figure*}

\paragraph{The abstraction can discard information the model needed.}
The \emph{abstract} harness keeps the diagnosis and drops the call. For error families whose repair is fully determined by the diagnosis, such as
an unknown tool or a missing required argument, that is lossless in the sense
that matters. For
families where the model needs to see \emph{what it wrote} in order to see what
was wrong with it, the abstraction can remove the evidence along with the
hazard. We expect this to bite hardest where the failure is about a specific
value rather than a structural mistake, and Section~\ref{sec:analysis}'s
per-error-family breakdown is where to look for it.

\paragraph{The effect weakens with scale, and the interventions with it.}
The inversion shrinks monotonically with model size within every family we
tested. It follows that the value of these interventions is largest exactly
where agents are cheapest to run and smallest where they are most capable. We do
not know where, or whether, the gain crosses zero; our ladder stops at
\LargestModel\ because that is what fits in memory in float32 on the hardware
available and finishes in the time available, and we
decline to extrapolate a crossing point from a fit whose zero lies outside the
range we measured. A reader with GPUs should treat ``does this vanish at 8B?''
as the first follow-up question, and we have made it cheap to answer: the probe
needs a few hundred forward passes per model, not rollouts.

\paragraph{The dominant failure at this scale is not repetition at all.}
It is worth being clear about what the agent gets wrong most often, because it
bounds how much any harness change can buy. Across rollouts, \ProseShare\ of actions are prose where a call was expected, for instance the
model announcing that it has cancelled the meeting instead of calling
\texttt{cancel\_event}. That is a
capability limit of a 0.5B model, and no arrangement of the transcript fixes it.
A further \TruncShare\ of actions are calls cut off by our \AgentTokenCap-token generation
budget, which is our artefact rather than the model's; it applies identically in
every condition and is small enough not to move the comparison, but it does
depress absolute success on the templates with long arguments.

The interventions we study act on repetition, and repetition is a substantial
but not majority share of what goes wrong here. That is the honest ceiling on
the effect sizes in Section~\ref{sec:agent}.

\paragraph{A regime where the standard harness is fine.}
When the failed action is one the model was unlikely to produce anyway, for
instance a call it emitted only because of a formatting accident, the copying
lift starts from a very low base and the absolute repeat probability stays
small. The inversion is a statement about a ratio; it is most dangerous when the
failed action was already plausible, which is precisely the case for the
near-miss errors that dominate real agent transcripts.

\section{Limitations}
\label{sec:limitations}

\paragraph{Scale.}
Our ladder stops at \LargestModel\ parameters. That is a hardware limit, not a
scientific choice: every model runs in float32 on a CPU with 17\,GB of RAM, and
a checkpoint that does not finish enough items in the time available is
excluded rather than reported thin (Appendix~\ref{app:repro}). The trend we
observe is monotone in size, so the honest summary is that we have characterised
the small-model regime and shown the direction of travel, not that we have
located a threshold. Whether frontier-scale models invert at all is untested
here.

\paragraph{Environments.}
ToolShed is our own construction. We built it because the central measurement
needs counterfactual observations that no public benchmark provides, and we have
tried to be explicit about the ways that choice could flatter us: the
perturbation operators are ours, the error messages are ours, and the task
templates are ours. CodeRepair is the check on that, with MBPP problems, real mutations, and
observations written by the Python interpreter rather than by us, and it shows
the same pattern. Neither is a long-horizon, realistic agent
benchmark. We would expect the effect to be larger, not smaller, in settings
with longer transcripts and more failures per trajectory, but we have not shown
that.

\paragraph{Perturbations are synthetic.}
Failing actions come from fixed operators rather than from model samples. This buys the property the scaling claim depends on, that every model is
probed on byte-identical items, at the cost of some realism. Sampling failures from each
model would make the item set model-dependent and would confound size with item
difficulty; we judged that the worse trade. The operators were chosen to match
the mistakes small models actually make, and the rollout study, where models
generate their own failures, is where that choice is put to the test.

\paragraph{Greedy decoding.}
Rollouts use greedy decoding, so each (model, harness) pair yields one
trajectory per task and our intervals over rollouts come from resampling tasks
rather than seeds. Sampling would add a variance component we have not measured.
Greedy is the right default for a comparison of harnesses, since it removes
decoding noise from a contrast we care about, but it is not how every deployed
agent runs.

\paragraph{Log-probabilities of whole strings.}
Corrective gain is a difference in summed token log-probability. Summed
quantities grow with string length, so the raw nat values are not comparable
across environments with very different action lengths. Within an environment
they are comparable across conditions and models because the string being scored
is identical; the normalised repeat probability and the exact greedy repeat rate
are provided precisely because they carry no such caveat.

\paragraph{A numerical failure worth knowing about.}
On this machine, calling \texttt{torch.set\_num\_threads($n$)} with $n > 1$
causes the installed PyTorch build to return all-NaN logits, silently, with
unchanged wall-clock time. We lost one complete probe run to this before
noticing, and our first verification script failed to catch it because
\texttt{max(0.0, nan)} returns \texttt{0.0} in Python, so a comparison of two
NaN quantities reported perfect agreement. Model loading now ends with a forward
pass that must produce finite logits, the scorer checks every returned logit
row, and the verification script rejects non-finite and positive
log-probabilities before comparing them. We record this because it is the kind
of fault that produces a clean-looking table of fabricated numbers, and because the mitigation, which is to assert on the numerical path and not
just on the code path, is cheap and general.

\section{Broader impact}

The interventions we study make small agents repeat themselves less, which
reduces wasted tool calls, wasted tokens and the wall-clock cost of a stuck
loop. The direct effects are mundane and mostly good. Two second-order points
are worth naming. First, a harness that suppresses repeated failed actions makes
an agent more persistent, and persistence is not always desirable: an agent that
keeps trying new calls after several failures may do more damage in a
side-effecting environment than one that gets stuck. Step budgets and
side-effect confirmation remain necessary; nothing here replaces them. Second,
our results make small models more usable as agents, which lowers the cost of
running autonomous systems locally. That is broadly positive for privacy and
access, and it also lowers the cost of running such systems without oversight.

\section{Conclusion}
\label{sec:conclusion}

Agent harnesses append a failed call and its error message to the transcript
because the error is information. It is. But the transcript entry that carries
it also carries the failed call's exact token sequence, and for small models
that second, unintended payload is much the larger of the two. Measured as a
change in the log-probability of writing the failed call again, the net effect
of the failure record is to make repetition \emph{more} likely, on every model
we tested, in both environments.

The decomposition is what makes this actionable rather than merely
discouraging. The harm is carried by the surface form, not by a failure to read
the message, so it is a property of the harness and not of the model, and harnesses are much
cheaper to change than models. Keeping the diagnosis while
withholding the string, or making the string unreachable at decoding time, acts
on the term that matters. Instructing the model not to repeat itself does not.

Neither does the remedy that sounds most obviously right. Deleting the failed
attempt and retrying from a clean context is the worst harness we measured,
because it restores precisely the context that produced the failure. The
principle that survives all three comparisons is narrow enough to be useful: the
context after a failure must differ from the context before it, and the
difference must not be the failed action.

Three questions follow directly. Does the inversion persist at frontier scale,
and if it disappears, does it do so because copying weakens or because the error
message finally wins? Our decomposition answers that question with a few hundred
forward passes per model, so it is cheap for anyone with the hardware. Second,
how should a harness decide \emph{what} to abstract? We used a fixed mapping from
error family to gloss; a runtime that knows more about its own failures could
say more, and there is a real trade-off between removing the hazard and removing
the evidence. Third, can the effect be trained away rather than engineered around, in the way
unlikelihood training and repetition penalties address open-ended text? Our results suggest what such a
training signal would have to target, and it is not the model's understanding of
error messages.

\bibliographystyle{plainnat}
\bibliography{references}

\clearpage
\appendix
\onecolumn
\section{Reproduction}
\label{app:repro}

\paragraph{Runs excluded for being incomplete.}
Every model/environment cell reported here reached at least 30 scored items; none were excluded on that ground.

\paragraph{Environment.}
Python 3.12, PyTorch \TorchVersion\ (CPU build), on \CPUName. No GPU is used or
required. Exact pinned versions are in \texttt{requirements.txt} and
\texttt{environment.yml}, and the software environment is recorded in
\texttt{results/raw/**/*.meta.json} for every run made after we started writing
it there.

The Transformers version is not single-valued: the environment was rebuilt
partway through the study, and runs exist under \TransformersVersions. Rather
than assert that this does not matter, we measured it.
\texttt{scripts/check\_version\_drift.py} re-scores stored rows in the current
environment and compares them to what is on disk; over \DriftChecked\
(item, condition, candidate) triples for \DriftModel\ the largest disagreement
is \DriftMax\ nats, roughly four orders of magnitude below the smallest effect
the paper reports and consistent with float32 non-determinism in the kernel
selection. Scoring is a deterministic forward pass, so this is the expected
result; we record it because the alternative is asking the reader to take it on
trust.

\paragraph{Pipeline.}
\begin{verbatim}
python scripts/download_models.py --ladder
python scripts/build_items.py
python scripts/verify_scoring.py --model smollm2-135m
python scripts/launch_probes.py --phases ABC
python scripts/run_agent_study.py --models qwen2.5-0.5b \
       --env toolshed --n-tasks 30 --max-steps 8
python scripts/analyze_probe.py
python scripts/analyze_agent.py
python scripts/make_facts.py && python scripts/make_figures.py
\end{verbatim}

Every stage is resumable. Probe results are keyed by
(item, condition, candidate) and existing keys are skipped, so an interrupted
run continues where it stopped; rollout files are written per
(model, environment, harness, temperature, seed).

\paragraph{Seeds.}
All randomness derives from base seed \texttt{20260808} through named
sub-streams (\texttt{slmecho.\allowbreak seeding.\allowbreak derive\_seed}), so
re-running one part of
the study reproduces the same items even if the surrounding loop changed. World
generation, task instantiation, item subsampling and the bootstrap each draw
from their own stream.

\paragraph{Generation parameters.}
Probe scoring is teacher-forced and involves no sampling. Rollouts use greedy
decoding with a \AgentTokenCap-token cap per action, stopping at the chat template's
end-of-turn token or at a newline (tool calls only; program repair allows
multi-line output with a 200-token cap).

\section{Probe conditions in full}
\label{app:conditions}

Each row is one way of writing the same decision point into a transcript. The
scored action string is identical across all of them; only the observation the
harness records differs.

\begin{table}[H]
\centering
\small
\begin{tabular}{l p{0.62\linewidth}}
\toprule
Condition & What the transcript contains \\
\midrule
\texttt{pre} & context before the attempt (baseline) \\
\texttt{fail} & attempt + error observation (standard ReAct harness) \\
\texttt{succ} & attempt + counterfactual success observation \\
\texttt{neut} & attempt + valence-free observation \\
\texttt{neut\_pad} & attempt + valence-free observation, length-matched to `fail` \\
\texttt{abstract} & harness-generated failure description, no verbatim call \\
\texttt{fail\_echo} & as `fail`, but the error quotes the failed call verbatim \\
\texttt{fail\_terse} & as `fail`, with the shortest error rendering \\
\texttt{fail\_verbose} & as `fail`, with the longest error rendering \\
\texttt{fail\_instr} & as `fail`, plus an explicit do-not-repeat instruction \\
\texttt{fail\_k2} & the same call fails twice \\
\texttt{fail\_k3} & the same call fails three times \\
\texttt{fail\_early} & as `fail`, but the failure precedes the successful steps \\
\texttt{fail\_other} & a DIFFERENT wrong action failed; the scored one was never written \\
\texttt{abstract\_min} & bare failure marker, no diagnosis and no verbatim call \\
\bottomrule
\end{tabular}
\caption{Every probe condition. All conditions for one item share the same system prompt, goal and successful prefix, byte for byte.}
\label{tab:conditions}
\end{table}

\section{Perturbation operators}
\label{app:operators}

Failing calls are not sampled from a model, they are constructed, so that the
failure is of a known kind and the correct call is known exactly. Each operator
takes a valid call and breaks it in one specific way. Examples are shown as
they appear in the transcript, truncated where they run past the column.

\begin{table}[H]
\centering
\small
\setlength{\tabcolsep}{3pt}
\begin{tabular}{l p{0.27\linewidth} p{0.27\linewidth} l}
\toprule
Operator & Reference call & Perturbed call & Error family \\
\midrule
\texttt{arg\_typo} & \texttt{\verbfont find\_\allowbreak{}contact(\allowbreak{}name=\allowbreak{}'Dana Bergmann')} & \texttt{\verbfont find\_\allowbreak{}contact(\allowbreak{}contact\_\allowbreak{}name=\allowbreak{}'Dana Bergmann')} & \texttt{\verbfont unexpected\_argument} \\
\texttt{date\_format} & \texttt{\verbfont list\_\allowbreak{}events(\allowbreak{}date=\allowbreak{}'2026-09-19')} & \texttt{\verbfont list\_\allowbreak{}events(\allowbreak{}date=\allowbreak{}'September 19,\allowbreak{} 2026')} & \texttt{\verbfont format\_error} \\
\texttt{drop\_arg} & \texttt{\verbfont send\_\allowbreak{}message(\allowbreak{}to=\allowbreak{}'dana.bergmann@corp.example',\allowbreak{} subject=\allowbreak{}'agenda',\allowbreak{} body=\allowbreak{}'roadmap,\allowbreak{} hiring,\allowbreak{} i\,\allowbreak{}\ldots{}} & \texttt{\verbfont send\_\allowbreak{}message(\allowbreak{}to=\allowbreak{}'dana.bergmann@corp.example',\allowbreak{} subject=\allowbreak{}'agenda')} & \texttt{\verbfont missing\_argument} \\
\texttt{hallucinated\_entity} & \texttt{\verbfont find\_\allowbreak{}contact(\allowbreak{}name=\allowbreak{}'Dana Bergmann')} & \texttt{\verbfont find\_\allowbreak{}contact(\allowbreak{}name=\allowbreak{}'Alex Ramirez')} & \texttt{\verbfont contact\_not\_found} \\
\texttt{name\_for\_email} & \texttt{\verbfont send\_\allowbreak{}message(\allowbreak{}to=\allowbreak{}'dana.bergmann@corp.example',\allowbreak{} subject=\allowbreak{}'agenda',\allowbreak{} body=\allowbreak{}'roadmap,\allowbreak{} hiring,\allowbreak{} i\,\allowbreak{}\ldots{}} & \texttt{\verbfont send\_\allowbreak{}message(\allowbreak{}to=\allowbreak{}'Dana Bergmann',\allowbreak{} subject=\allowbreak{}'agenda',\allowbreak{} body=\allowbreak{}'roadmap,\allowbreak{} hiring,\allowbreak{} incident revie\,\allowbreak{}\ldots{}} & \texttt{\verbfont format\_error} \\
\texttt{readonly\_target} & \texttt{\verbfont write\_\allowbreak{}file(\allowbreak{}path=\allowbreak{}'reports/roster.md',\allowbreak{} content=\allowbreak{}'Elena Okafor,\allowbreak{} Karim Okafor')} & \texttt{\verbfont write\_\allowbreak{}file(\allowbreak{}path=\allowbreak{}'system/config.ini',\allowbreak{} content=\allowbreak{}'Elena Okafor,\allowbreak{} Karim Okafor')} & \texttt{\verbfont permission\_error} \\
\texttt{tool\_typo} & \texttt{\verbfont find\_\allowbreak{}contact(\allowbreak{}name=\allowbreak{}'Dana Bergmann')} & \texttt{\verbfont find\_\allowbreak{}contacts(\allowbreak{}name=\allowbreak{}'Dana Bergmann')} & \texttt{\verbfont unknown\_tool} \\
\texttt{type\_swap} & \texttt{\verbfont create\_\allowbreak{}event(\allowbreak{}title=\allowbreak{}'1:1',\allowbreak{} date=\allowbreak{}'2026-10-13',\allowbreak{} time=\allowbreak{}'09:30',\allowbreak{} attendees=\allowbreak{}['karim.silva@corp.\,\allowbreak{}\ldots{}} & \texttt{\verbfont create\_\allowbreak{}event(\allowbreak{}title=\allowbreak{}'1:1',\allowbreak{} date=\allowbreak{}'2026-10-13',\allowbreak{} time=\allowbreak{}'09:30',\allowbreak{} attendees=\allowbreak{}'karim.silva@corp.e\,\allowbreak{}\ldots{}} & \texttt{\verbfont type\_error} \\
\bottomrule
\end{tabular}
\caption{Perturbation operators used to build failing actions in ToolShed, with one instantiated example each.}
\label{tab:operators}
\end{table}

\section{Per-error-family results}
\label{app:by-error}

The same headline quantity, split by the kind of error the failing call
provoked. The point of the table is the absence of structure: if the effect
were an artefact of one error message's wording, one row would carry it.


\begin{table}[H]
\centering
\tblfontsmall
\setlength{\tabcolsep}{\tblsepsmall}
\begin{tabular}{l c c c c c c}
\toprule
Error family & SmolLM2-135M & SmolLM2-360M & Qwen2.5-0.5B & Qwen3-0.6B & Llama-3.2-1B & SmolLM2-1.7B \\
\midrule
\texttt{permission\_error} & \textbf{\textminus 44.4\,\ci{\textminus 47.0}{\textminus 42.2}} & \textbf{\textminus 43.0\,\ci{\textminus 44.9}{\textminus 41.3}} & \textbf{\textminus 37.9\,\ci{\textminus 42.4}{\textminus 33.9}} & \textbf{\textminus 51.9\,\ci{\textminus 52.9}{\textminus 51.1}} & -- & \textbf{\textminus 30.1\,\ci{\textminus 33.1}{\textminus 27.4}} \\
\texttt{contact\_not\_found} & \textbf{\textminus 34.4\,\ci{\textminus 36.4}{\textminus 31.3}} & \textbf{\textminus 21.9\,\ci{\textminus 23.0}{\textminus 20.5}} & \textbf{\textminus 27.8\,\ci{\textminus 30.9}{\textminus 26.1}} & -- & -- & \textbf{\textminus 19.0\,\ci{\textminus 21.3}{\textminus 17.5}} \\
\texttt{format\_error} & \textbf{\textminus 22.9\,\ci{\textminus 24.5}{\textminus 21.1}} & \textbf{\textminus 18.4\,\ci{\textminus 20.1}{\textminus 16.9}} & \textbf{\textminus 15.1\,\ci{\textminus 17.8}{\textminus 12.4}} & \textbf{\textminus 23.9\,\ci{\textminus 27.5}{\textminus 19.9}} & \textbf{\textminus 14.7\,\ci{\textminus 18.2}{\textminus 11.6}} & \textbf{\textminus 13.2\,\ci{\textminus 14.5}{\textminus 12.1}} \\
\texttt{file\_not\_found} & -- & -- & -- & -- & -- & \textbf{\textminus 16.9\,\ci{\textminus 18.7}{\textminus 14.1}} \\
\texttt{event\_not\_found} & \textbf{\textminus 18.7\,\ci{\textminus 19.2}{\textminus 18.4}} & \textbf{\textminus 13.0\,\ci{\textminus 13.7}{\textminus 12.3}} & \textbf{\textminus 11.7\,\ci{\textminus 12.4}{\textminus 11.2}} & -- & -- & -- \\
\texttt{missing\_argument} & \textbf{\textminus 17.3\,\ci{\textminus 20.2}{\textminus 13.6}} & \textbf{\textminus 15.4\,\ci{\textminus 18.6}{\textminus 11.5}} & \textbf{\textminus 10.4\,\ci{\textminus 12.0}{\textminus 8.3}} & \textbf{\textminus 22.7\,\ci{\textminus 24.7}{\textminus 19.3}} & \textbf{\textminus 9.0\,\ci{\textminus 15.0}{\textminus 1.1}} & \textbf{\textminus 9.8\,\ci{\textminus 12.1}{\textminus 7.2}} \\
\texttt{unknown\_tool} & \textbf{\textminus 18.6\,\ci{\textminus 22.4}{\textminus 15.4}} & \textbf{\textminus 13.5\,\ci{\textminus 18.9}{\textminus 9.0}} & \textbf{\textminus 9.8\,\ci{\textminus 12.3}{\textminus 7.4}} & \textbf{\textminus 16.1\,\ci{\textminus 19.9}{\textminus 13.2}} & \textbf{\textminus 11.0\,\ci{\textminus 15.4}{\textminus 7.5}} & \textbf{\textminus 12.2\,\ci{\textminus 14.6}{\textminus 9.9}} \\
\texttt{attendee\_not\_found} & -- & -- & -- & -- & -- & \textbf{\textminus 12.3\,\ci{\textminus 12.9}{\textminus 11.1}} \\
\texttt{unexpected\_argument} & \textbf{\textminus 19.0\,\ci{\textminus 22.9}{\textminus 14.8}} & \textbf{\textminus 12.3\,\ci{\textminus 16.4}{\textminus 8.8}} & \textbf{\textminus 5.9\,\ci{\textminus 10.2}{\textminus 1.1}} & \textbf{\textminus 9.1\,\ci{\textminus 13.4}{\textminus 5.5}} & \textbf{\textminus 8.6\,\ci{\textminus 11.7}{\textminus 5.8}} & \textbf{\textminus 10.1\,\ci{\textminus 11.4}{\textminus 8.8}} \\
\texttt{type\_error} & \textbf{\textminus 18.9\,\ci{\textminus 19.6}{\textminus 18.2}} & \textbf{\textminus 11.2\,\ci{\textminus 12.6}{\textminus 9.8}} & \textminus 0.7\,\ci{\textminus 3.2}{1.8} & \textbf{\textminus 6.3\,\ci{\textminus 11.8}{\textminus 1.3}} & \textbf{\textminus 3.2\,\ci{\textminus 4.0}{\textminus 2.3}} & \textbf{\textminus 7.7\,\ci{\textminus 10.2}{\textminus 5.3}} \\
\bottomrule
\end{tabular}
\caption{Corrective gain (nats) by error family on ToolShed, sorted by pooled mean. Rows are the error the failing call provoked; a broadly uniform column argues against any explanation resting on the wording of one error type.}
\label{tab:by-error}
\end{table}

\section{Repetition statistics on program repair}
\label{app:repeat-code}

The main text reports the repetition readouts for the simulated tool
environment; the same quantities for MBPP program repair are below. They are
computed identically, over the same fixed candidate set construction.


\begin{table*}[t]
\centering
\tblfont
\setlength{\tabcolsep}{\tblsep}
\begin{tabular}{l r c c c c c c}
\toprule
Model & Params & $p_{\text{rep}}$ before & $p_{\text{rep}}$ after & $\Delta p_{\text{rep}}$ & greedy before & greedy after & greedy abs. \\
\midrule
SmolLM2-135M & 0.14B & \textbf{0.111\,\ci{0.053}{0.180}} & \textbf{0.951\,\ci{0.917}{0.978}} & \textbf{0.840\,\ci{0.770}{0.901}} & 0.000\,\ci{0.000}{0.000} & 0.000\,\ci{0.000}{0.000} & 0.000\,\ci{0.000}{0.000} \\
SmolLM2-360M & 0.36B & \textbf{0.089\,\ci{0.032}{0.156}} & \textbf{0.856\,\ci{0.795}{0.910}} & \textbf{0.767\,\ci{0.689}{0.840}} & 0.000\,\ci{0.000}{0.000} & \textbf{0.100\,\ci{0.033}{0.183}} & 0.000\,\ci{0.000}{0.000} \\
Qwen2.5-0.5B & 0.49B & \textbf{0.075\,\ci{0.023}{0.138}} & \textbf{0.833\,\ci{0.766}{0.893}} & \textbf{0.758\,\ci{0.680}{0.831}} & 0.000\,\ci{0.000}{0.000} & 0.017\,\ci{0.000}{0.050} & 0.000\,\ci{0.000}{0.000} \\
\bottomrule
\end{tabular}
\caption{Repetition probabilities on coderepair. $p_{\text{rep}}$ is the normalised probability of the failed call among the scored candidate actions; \emph{greedy} is the fraction of items on which greedy decoding reproduces the failed call exactly.}
\label{tab:probe-repeat-coderepair}
\end{table*}

\section{Prompts}
\label{app:prompts}

The complete system prompts, the failure-abstraction template, and the
``do not repeat'' instruction are reproduced verbatim below; they are also
emitted by \texttt{scripts/dump\_prompts.py} so they cannot drift from the code.
Two typesetting conventions apply and nothing else has been altered: a line too
long for the page is broken and the broken line ends in a backslash, and the
few non-ASCII characters the prompts contain (an em dash, curly quotes) are
transliterated, since the typewriter font has no glyph for them.

\paragraph{ToolShed system prompt (probe)}
{\verbfont\begin{verbatim}
You are a tool-using assistant working inside a small office workspace. Solve the user's request by \
calling tools one at a time.

Rules:
- Reply with exactly one tool call and nothing else.
- Use the form tool_name(arg=value), with keyword arguments.
- Dates are YYYY-MM-DD and times are HH:MM.
- When the request is fully satisfied, reply with finish().

Available tools:
- list_files(folder: str) -> list[str]
    List the files inside a folder.
    folder: folder name, e.g. 'notes'
- read_file(path: str) -> str
    Return the contents of a file.
    path: full path, e.g. 'notes/todo.md'
- write_file(path: str, content: str) -> str
    Create or overwrite a file.
    path: full path
    content: text to store
- delete_file(path: str) -> str
    Delete a file.
    path: full path
- find_contact(name: str) -> dict
    Look up one contact by full name.
    name: exact full name
- list_contacts(team: str) -> list[dict]
    List the members of a team.
    team: team name, e.g. 'design'
- list_events(date: str) -> list[dict]
    List calendar events on a date.
    date: date as YYYY-MM-DD
- create_event(title: str, date: str, time: str, attendees: list[str]) -> str
    Add a calendar event.
    title: event title
    date: date as YYYY-MM-DD
    time: start time as HH:MM
    attendees: list of attendee emails
- cancel_event(event_id: str) -> str
    Cancel a calendar event by id.
    event_id: event id, e.g. 'evt-4'
- send_message(to: str, subject: str, body: str) -> str
    Send a message to one recipient email address.
    to: recipient email address
    subject: subject line
    body: message body
- set_reminder(text: str, date: str) -> str
    Store a reminder for a date.
    text: reminder text
    date: date as YYYY-MM-DD
- convert_units(value: float, from_unit: str, to_unit: str) -> float
    Convert a value between two units.
    value: numeric value
    from_unit: source unit
    to_unit: target unit
\end{verbatim}}

\paragraph{Format demonstration appended in the agent study}
{\verbfont\begin{verbatim}

A valid reply looks exactly like this, with no prefix or explanation:
list_files(folder='notes')
\end{verbatim}}

\paragraph{CodeRepair system prompt}
{\verbfont\begin{verbatim}
You are a Python programmer. The user gives a task and the tests that will be run against your code.

Rules:
- Reply with one Python function definition and nothing else.
- Do not include the tests, comments, explanations or markdown fences.
- Use exactly the function name that appears in the tests.
\end{verbatim}}

\paragraph{Do-not-repeat instruction}
{\verbfont\begin{verbatim}
Important: some of your earlier tool calls failed. Do not repeat a call that has already failed; change \
the call before trying again.
\end{verbatim}}

\paragraph{Failure abstraction (example)}
{\verbfont\begin{verbatim}
[attempt 1 failed: find_contact --- an argument name was not recognised; that call is not repeatable]
\end{verbatim}}

\paragraph{Error-family glosses used by the abstraction}
{\verbfont\begin{verbatim}
arity_error              -> wrong number of arguments
attendee_not_found       -> no such attendee
contact_not_found        -> no such contact
event_not_found          -> no such event
file_not_found           -> no such file
folder_not_found         -> no such folder
format_error             -> an argument was badly formatted
missing_argument         -> a required argument was missing
parse_error              -> the call could not be parsed
permission_error         -> the target was read-only
recipient_not_found      -> no such recipient
runtime_error            -> the code raised an exception
state_error              -> the target was already in the requested state
syntax_error             -> the code did not compile
team_not_found           -> no such team
test_failure             -> the code did not pass the tests
type_error               -> an argument had the wrong type
unexpected_argument      -> an argument name was not recognised
unknown_tool             -> no such tool
unsupported_conversion   -> that conversion is not supported
\end{verbatim}}

\section{Example trajectories}
\label{app:trajectories}

One task under every harness, chosen by rule: the task with the most exact
repeats under the standard harness. It is a single task and it is the worst
one, so it illustrates rather than demonstrates, and the quantitative claims are in
Section~\ref{sec:agent}, but it is worth reading for what it shows and
also for what it complicates.

Under \emph{verbatim}, \emph{drop} and \emph{abstract} the model writes the
same call six times. Under the two ban harnesses it cannot, and what it
produces instead is the same call with material accreted onto it: fragments of
the error message under \emph{verbatim+ban}, chained
\texttt{.replace} calls under \emph{abstract+ban}. Neither is a reconsidered
call. \emph{verbatim+instr} is the one that behaves differently here, trying
four distinct calls, none of them right.

Two of those cut against the aggregates and we would rather show that than
pick a friendlier task. The abstraction reduces repetition across the task set
but does nothing on this item, and the instruction does nothing across the task
set but changes the trajectory on this one. Task-level variance at this scale
is large, which is why the paper's claims rest on paired comparisons over all
tasks rather than on examples.

\textbf{qwen2.5-0.5b, harness=verbatim}
{\verbfont\begin{verbatim}
goal: agent-0011  solved=False
1. [ok ] convert_units(value=88, from_unit='km', to_unit='mi')
2. [ERR] write_file(path='reports/conversion.md', content=str(54.6806))
3. [ERR] write_file(path='reports/conversion.md', content=str(54.6806))
4. [ERR] write_file(path='reports/conversion.md', content=str(54.6806))
5. [ERR] write_file(path='reports/conversion.md', content=str(54.6806))
6. [ERR] write_file(path='reports/conversion.md', content=str(54.6806))
\end{verbatim}}

\textbf{qwen2.5-0.5b, harness=verbatim-instr}
{\verbfont\begin{verbatim}
goal: agent-0011  solved=False
1. [ok ] convert_units(value=88, from_unit='km', to_unit='mi')
2. [ERR] read_file(path='observations/54.6806.txt')
3. [ERR] delete_file(path='observations/54.6806.txt')
4. [ERR] list_files(folder='observations')
5. [ok ] list_files(folder='drafts')
6. [ok ] list_files(folder='drafts')
\end{verbatim}}

\textbf{qwen2.5-0.5b, harness=drop}
{\verbfont\begin{verbatim}
goal: agent-0011  solved=False
1. [ok ] convert_units(value=88, from_unit='km', to_unit='mi')
2. [ERR] write_file(path='reports/conversion.md', content=str(54.6806))
3. [ERR] write_file(path='reports/conversion.md', content=str(54.6806))
4. [ERR] write_file(path='reports/conversion.md', content=str(54.6806))
5. [ERR] write_file(path='reports/conversion.md', content=str(54.6806))
6. [ERR] write_file(path='reports/conversion.md', content=str(54.6806))
\end{verbatim}}

\textbf{qwen2.5-0.5b, harness=abstract}
{\verbfont\begin{verbatim}
goal: agent-0011  solved=False
1. [ok ] convert_units(value=88, from_unit='km', to_unit='mi')
2. [ERR] write_file(path='reports/conversion.md', content=str(54.6806))
3. [ERR] write_file(path='reports/conversion.md', content=str(54.6806))
4. [ERR] write_file(path='reports/conversion.md', content=str(54.6806))
5. [ERR] write_file(path='reports/conversion.md', content=str(54.6806))
6. [ERR] write_file(path='reports/conversion.md', content=str(54.6806))
\end{verbatim}}

\textbf{qwen2.5-0.5b, harness=verbatim-ban}
{\verbfont\begin{verbatim}
goal: agent-0011  solved=False
1. [ok ] convert_units(value=88, from_unit='km', to_unit='mi')
2. [ERR] write_file(path='reports/conversion.md', content=str(54.6806))
3. [ERR] write_file(path='reports/conversion.md', content=str(54.6806), error='write_file: argument value ~
4. [ERR] write_file(path='reports/conversion.md', content=str(54.6806), error='write_file: SyntaxError: co~
5. [ERR] write_file(path='reports/conversion.md', content=str(54.6806), error='write_file: SyntaxError: co~
6. [ERR] write_file(path='reports/conversion.md', content=str(54.6806), error='write_file: argument value ~
\end{verbatim}}

\textbf{qwen2.5-0.5b, harness=abstract-ban}
{\verbfont\begin{verbatim}
goal: agent-0011  solved=False
1. [ok ] convert_units(value=88, from_unit='km', to_unit='mi')
2. [ERR] write_file(path='reports/conversion.md', content=str(54.6806))
3. [ERR] write_file(path='reports/conversion.md', content=str(54.6806).replace('.', ','))
4. [ERR] write_file(path='reports/conversion.md', content=str(54.6806).replace('.', ',').replace('-', ','))
5. [ERR] write_file(path='reports/conversion.md', content=str(54.6806).replace('.', ',').replace('-', ','))
6. [ERR] write_file(path='reports/conversion.md', content=str(54.6806).replace('.', ',').replace('-', ','))
\end{verbatim}}

\end{document}